%% file: DESY-10-129.tex
\documentclass[zpreprint,zbstdefault]{zeus_paper}
\usepackage[english]{babel}

\input zeus_def.tex
\input def.tex
\input  DESY-10-129-cit.tex
\begin{document}
\include{DESY-10-129-tit}

\include{DESY-10-129-aut}
\include{DESY-10-129-txt}
\include{DESY-10-129-ref}

\include{DESY-10-129-tab}

\include{DESY-10-129-fig}
%
%
\end{document}

%% file: zeus_def.tex
\newcommand{\ZcoosysB}{%
The ZEUS coordinate system is a right-handed Cartesian system, with the $Z$
axis pointing in the proton beam direction, referred to as the ``forward
direction'', and the $X$ axis pointing towards the centre of HERA.
The coordinate origin is at the nominal interaction point.\xspace}

\newcommand{\ZcoosysfnB}{\footnote{\ZcoosysB}}

\chardef\usc=95
\chardef\til=126
\catcode`\@=11 
\DeclareRobustCommand\xdotspace{\futurelet\@let@token\@xdotspace}
\def\@xdotspace{%
  \ifx\@let@token.\else
  \ifx\@let@token\bgroup.\else
  \ifx\@let@token\egroup.\else
  \ifx\@let@token\/.\else
  \ifx\@let@token\ .\else
  \ifx\@let@token~.\else
  \ifx\@let@token!.\else
  \ifx\@let@token,.\else
  \ifx\@let@token:.\else
  \ifx\@let@token;.\else
  \ifx\@let@token?.\else
  \ifx\@let@token/.\else
  \ifx\@let@token'.\else
  \ifx\@let@token).\else
  \ifx\@let@token-.\else
  \ifx\@let@token\@xobeysp.\else
  \ifx\@let@token\space.\else
  \ifx\@let@token\@sptoken.\else
   .\space
   \fi\fi\fi\fi\fi\fi\fi\fi\fi\fi\fi\fi\fi\fi\fi\fi\fi\fi}
\catcode`\@=12 

\newcommand{\stru}[2]{%
   \relax\ifmmode\hbox{\vrule height#1 depth#2 width0pt}%
   \else\vrule height#1 depth#2 width0pt\fi}

\newcommand{\Ronum}[1]{\uppercase\expandafter{\romannumeral#1}}
\newcommand{\ronum}[1]{\expandafter{\romannumeral#1}}
\DeclareRobustCommand{\LaTeXZ}{%
  \LaTeX\kern-.05em4\kern-.1em
  {\raisebox{-0.2ex}{$\scriptstyle\text{ZEUS}$}}\xspace}

\DeclareMathAlphabet{\mathbf}{OT1}{cmr}{bx}{sl}
\newcommand{\eVdist}{\kern-0.06667em}

\newcommand{\Gev}{{\text{Ge}\eVdist\text{V\/}}}

\newcommand{\gev}{{\,\text{Ge}\eVdist\text{V\/}}}

\newcommand{\pb}{\,\text{pb}}

\newcommand{\Tesla}{\,\text{T}}

\newcommand{\slashfrac}[2]{%
  \raisebox{0.5ex}{\ensuremath #1}\kern-0.12em/\kern-0.08em
  \raisebox{-.8ex}{\ensuremath #2}}

\newcommand{\sqr}[3]{%
    {\vcenter{\hrule height.#3ex\hbox{\vrule width.#2ex height#1ex
     \kern#1ex\vrule width.#3ex}\hrule height.#2ex}}}

\catcode`\@=11 
\newcommand{\parenbar}{\mathpalette\p@renb@r}
\def\p@renb@r#1#2{\vbox{%
  \ifx#1\scriptscriptstyle \dimen@.7em\dimen@ii.2em\else
  \ifx#1\scriptstyle \dimen@.8em\dimen@ii.25em\else
  \dimen@1em\dimen@ii.4em\fi\fi \offinterlineskip
  \ialign{\hfill##\hfill\cr
    \vbox{\hrule width\dimen@ii}\cr
    \noalign{\vskip-.3ex}%
    \hbox to\dimen@{$\mathchar300\hfil\mathchar301$}\cr
    \noalign{\vskip-.3ex}%
    $#1#2$\cr}}}
\catcode`\@=12 

\newcommand{\MSbar}{\hbox{$\overline{\rm MS}$}\xspace}

\newcommand{\rnge}{\hbox{$\,\text{--}\,$}}

\newcommand{\IP}{{\rm I$\kern-0.01667em$P}\xspace}

\mathchardef\qsm=63
\mathchardef\pls=43
\mathchardef\mns=512
\mathchardef\plm=518
\mathchardef\eql=61
\mathchardef\smallleft=300
\mathchardef\smallright=301
\mathchardef\les=316
\mathchardef\gre=318
\mathchardef\leq=532
\mathchardef\grq=533

\catcode`\@=11 
\newcounter{pict@width}
\newcounter{pict@height}
\newlength{\pict@scale}
\newcommand{\psfigadd}[4]{%
\setcounter{pict@width}{1*\ratio{#2+\pict@scale/2}{\pict@scale}}
\setcounter{pict@height}{1*\ratio{#3+\pict@scale/2}{\pict@scale}}
\setlength{\unitlength}{\pict@scale}
\hbox to #2{\hspace{-\fill}\begin{picture}(\thepict@width,\thepict@height)
\put(0,0){\psfig{figure=#1,width=#2,height=#3,clip=}}
\SetScale{0.283466457}
\SetWidth{1.763889}
{#4}
\end{picture}}
}
\newcounter{pict@widthfst}
\newcounter{pict@widthscd}
\newcounter{pict@widthtot}
\newcommand{\psfigaddtwo}[7]{%
\setcounter{pict@widthfst}{1*\ratio{#2+\pict@scale/2}{\pict@scale}}
\setcounter{pict@widthscd}{1*\ratio{#2+#4+\pict@scale/2}{\pict@scale}}
\setcounter{pict@widthtot}{1*\ratio{#2+#4+#6+\pict@scale/2}{\pict@scale}}
\setcounter{pict@height}{1*\ratio{#3+\pict@scale/2}{\pict@scale}}
\setlength{\unitlength}{\pict@scale}
\hbox{\hspace{-\fill}\begin{picture}(\thepict@widthtot,\thepict@height)
\put(0,0){\psfig{figure=#1,width=#2,height=#3,clip=}}
\put(\thepict@widthscd,0){\psfig{figure=#5,width=#6,height=#3,clip=}}
\SetScale{0.283466457}
\SetWidth{1.763889}
{#7}
\end{picture}}
}
\newcommand{\psfigror}[4]{%
\setcounter{pict@width}{1*\ratio{#2+\pict@scale/2}{\pict@scale}}
\setcounter{pict@height}{1*\ratio{#3+\pict@scale/2}{\pict@scale}}
\setlength{\unitlength}{\pict@scale}
\hbox{\begin{picture}(\thepict@width,\thepict@height)
\put(0,\thepict@height){\psfig{figure=#1,width=#3,height=#2,clip=,angle=270}}
\SetScale{0.283466457}
\SetWidth{1.763889}
{#4}
\end{picture}}
}
\newcommand{\psfigrol}[4]{%
\setcounter{pict@width}{1*\ratio{#2+\pict@scale/2}{\pict@scale}}
\setcounter{pict@height}{1*\ratio{#3+\pict@scale/2}{\pict@scale}}
\setlength{\unitlength}{\pict@scale}
\hbox{\begin{picture}(\thepict@width,\thepict@height)
\put(0,0){\psfig{figure=#1,width=#3,height=#2,clip=,angle=90}}
\SetScale{0.283466457}
\SetWidth{1.763889}
{#4}
\end{picture}}
}
\catcode`\@=12 
\newlength\listtextwidth

\catcode`\@=11 
\newlength{\@tabfninsert}
\newlength{\@tabfnwidth}
\newcommand{\tabfootnote}[2]{%
  \setlength{\@tabfninsert}{0.8em}
  \setlength{\@tabfnwidth}{\textwidth}
  \addtolength{\@tabfnwidth}{-\@tabfninsert}
  \addtolength{\@tabfnwidth}{-0.4em}
  \noindent\makebox[\@tabfninsert][r]{\footnotesize$^{#1}$\hfil}\hfill%
  \parbox[t]{\@tabfnwidth}{\footnotesize #2\hfill}}
\catcode`\@=12 

%% file: def.tex
\newcommand{\PTM}       {P_{T,{\rm miss}}}
\newcommand{\PTvector}     {\overrightarrow{P}_{T}}
\newcommand{\PTMvector}    {\overrightarrow{P}_{T,{\rm miss}}}

%% file: DESY-10-129-cit.tex
\def\citeCTD{{\cite{%
nim:a279:290,*npps:b32:181,*nim:a338:254%
}}\xspace}
\def\citeMVD{{\cite{%
nim:a581:656%
}}\xspace}
\def\citeCAL{{\cite{%
nim:a309:77,*nim:a309:101,*nim:a321:356,*nim:a336:23%
}}\xspace}

%% file: DESY-10-129-tit.tex
\title{Measurement of high-$\bold{Q^2}$ charged current deep inelastic scattering cross sections with a longitudinally polarised positron beam at HERA }                                                       
\author{ZEUS Collaboration}
\prepnum{DESY-10-129}
\draftversion{0.8, Post-Reading}
\date{August 2010}

\abstract{
Measurements of the cross sections for charged current  deep inelastic
scattering in $e^{+}p$ collisions with a  longitudinally polarised
positron beam are presented. The measurements are based on a data
sample with an integrated luminosity  of $132~\pb^{-1}$ collected
with the ZEUS detector at HERA at a centre-of-mass
energy of 318~\gev. The total cross section is presented at positive
and negative  values of the longitudinal polarisation of the positron
beams. The single-differential cross-sections $d\sigma/dQ^{2}$,
$d\sigma/dx$ and $d\sigma/dy$ are presented
for \mbox{$Q^{2}>200\gev^2$}. The reduced cross-section
$\tilde{\sigma}$ is presented  in the kinematic
range \mbox{$200<Q^2<60\,000 \gev^2$} and \mbox{$0.006<x<0.562$}.
The measurements agree well with the predictions of  the Standard
Model. The results are used to determine a lower limit on the mass 
of a hypothetical right-handed $W$ boson.}


\makezeustitle

%% file: DESY-10-129-aut.tex
%
%
%
%


\def\3{\ss}
\pagenumbering{Roman}
                                                   %
\begin{center}
{                      \Large  The ZEUS Collaboration              }
\end{center}

{\small


{\mbox H.~Abramowicz$^{44, ag}$, }
{\mbox I.~Abt$^{34}$, }
{\mbox L.~Adamczyk$^{13}$, }
{\mbox M.~Adamus$^{53}$, }
{\mbox R.~Aggarwal$^{7}$, }
{\mbox S.~Antonelli$^{4}$, }
{\mbox P.~Antonioli$^{3}$, }
{\mbox A.~Antonov$^{32}$, }
{\mbox M.~Arneodo$^{49}$, }
{\mbox V.~Aushev$^{26, ab}$
% ...
}
\newpage


\makebox[3em]{$^{1}$}
\begin{minipage}[t]{14cm}
{\it Argonne National Laboratory, Argonne, Illinois 60439-4815, USA}~$^{A}$

\end{minipage}\\
\makebox[3em]{$^{2}$}
\begin{minipage}[t]{14cm}
{\it Andrews University, Berrien Springs, Michigan 49104-0380, USA}

\end{minipage}\\
\makebox[3em]{$^{3}$}
\begin{minipage}[t]{14cm}
{\it INFN Bologna, Bologna, Italy}~$^{B}$

\end{minipage}\\
\makebox[3em]{$^{4}$}
\begin{minipage}[t]{14cm}
{\it University and INFN Bologna, Bologna, Italy}~$^{B}$

\end{minipage}\\
\makebox[3em]{$^{5}$}
\begin{minipage}[t]{14cm}
{\it Physikalisches Institut der Universit\"at Bonn,
Bonn, Germany}~$^{C}$

\end{minipage}\\
\makebox[3em]{$^{6}$}
\begin{minipage}[t]{14cm}
{\it H.H.~Wills Physics Laboratory, University of Bristol,
Bristol, United Kingdom}~$^{D}$

\end{minipage}\\
\makebox[3em]{$^{7}$}
\begin{minipage}[t]{14cm}
{\it Panjab University, Department of Physics, Chandigarh, India}

\end{minipage}\\
\makebox[3em]{$^{8}$}
\begin{minipage}[t]{14cm}
{\it Calabria University,
Physics Department and INFN, Cosenza, Italy}~$^{B}$

\end{minipage}\\
\makebox[3em]{$^{9}$}
\begin{minipage}[t]{14cm}
{\it Institute for Universe and Elementary Particles, Chonnam National University,\\
Kwangju, South Korea}

\end{minipage}\\
\makebox[3em]{$^{10}$}
\begin{minipage}[t]{14cm}
{\it Jabatan Fizik, Universiti Malaya, 50603 Kuala Lumpur, Malaysia}~$^{E}$

\end{minipage}\\
\makebox[3em]{$^{11}$}
\begin{minipage}[t]{14cm}
{\it Nevis Laboratories, Columbia University, Irvington on Hudson,
New York 10027, USA}~$^{F}$

\end{minipage}\\
\makebox[3em]{$^{12}$}
\begin{minipage}[t]{14cm}
{\it The Henryk Niewodniczanski Institute of Nuclear Physics, Polish Academy of \\
Sciences, Cracow, Poland}~$^{G}$

\end{minipage}\\
\makebox[3em]{$^{13}$}
\begin{minipage}[t]{14cm}
{\it Faculty of Physics and Applied Computer Science, AGH-University of Science and \\
Technology, Cracow, Poland}~$^{H}$

\end{minipage}\\
\makebox[3em]{$^{14}$}
\begin{minipage}[t]{14cm}
{\it Department of Physics, Jagellonian University, Cracow, Poland}

\end{minipage}\\
\makebox[3em]{$^{15}$}
\begin{minipage}[t]{14cm}
{\it Deutsches Elektronen-Synchrotron DESY, Hamburg, Germany}

\end{minipage}\\
\makebox[3em]{$^{16}$}
\begin{minipage}[t]{14cm}
{\it Deutsches Elektronen-Synchrotron DESY, Zeuthen, Germany}

\end{minipage}\\
\makebox[3em]{$^{17}$}
\begin{minipage}[t]{14cm}
{\it INFN Florence, Florence, Italy}~$^{B}$

\end{minipage}\\
\makebox[3em]{$^{18}$}
\begin{minipage}[t]{14cm}
{\it University and INFN Florence, Florence, Italy}~$^{B}$

\end{minipage}\\
\makebox[3em]{$^{19}$}
\begin{minipage}[t]{14cm}
{\it Fakult\"at f\"ur Physik der Universit\"at Freiburg i.Br.,
Freiburg i.Br., Germany}

\end{minipage}\\
\makebox[3em]{$^{20}$}
\begin{minipage}[t]{14cm}
{\it School of Physics and Astronomy, University of Glasgow,
Glasgow, United Kingdom}~$^{D}$

\end{minipage}\\
\makebox[3em]{$^{21}$}
\begin{minipage}[t]{14cm}
{\it Department of Engineering in Management and Finance, Univ. of
the Aegean, Chios, Greece}

\end{minipage}\\
\makebox[3em]{$^{22}$}
\begin{minipage}[t]{14cm}
{\it Hamburg University, Institute of Experimental Physics, Hamburg,
Germany}~$^{I}$

\end{minipage}\\
\makebox[3em]{$^{23}$}
\begin{minipage}[t]{14cm}
{\it Imperial College London, High Energy Nuclear Physics Group,
London, United Kingdom}~$^{D}$

\end{minipage}\\
\makebox[3em]{$^{24}$}
\begin{minipage}[t]{14cm}
{\it Institute of Particle and Nuclear Studies, KEK,
Tsukuba, Japan}~$^{J}$

\end{minipage}\\
\makebox[3em]{$^{25}$}
\begin{minipage}[t]{14cm}
{\it Institute of Physics and Technology of Ministry of Education and
Science of Kazakhstan, Almaty, Kazakhstan}

\end{minipage}\\
\makebox[3em]{$^{26}$}
\begin{minipage}[t]{14cm}
{\it Institute for Nuclear Research, National Academy of Sciences, and
Kiev National University, Kiev, Ukraine}

\end{minipage}\\
\makebox[3em]{$^{27}$}
\begin{minipage}[t]{14cm}
{\it Kyungpook National University, Center for High Energy Physics, Daegu,
South Korea}~$^{K}$

\end{minipage}\\
\makebox[3em]{$^{28}$}
\begin{minipage}[t]{14cm}
{\it Institut de Physique Nucl\'{e}aire, Universit\'{e} Catholique de Louvain, Louvain-la-Neuve,\\
Belgium}~$^{L}$

\end{minipage}\\
\makebox[3em]{$^{29}$}
\begin{minipage}[t]{14cm}
{\it Departamento de F\'{\i}sica Te\'orica, Universidad Aut\'onoma
de Madrid, Madrid, Spain}~$^{M}$

\end{minipage}\\
\makebox[3em]{$^{30}$}
\begin{minipage}[t]{14cm}
{\it Department of Physics, McGill University,
Montr\'eal, Qu\'ebec, Canada H3A 2T8}~$^{N}$

\end{minipage}\\
\makebox[3em]{$^{31}$}
\begin{minipage}[t]{14cm}
{\it Meiji Gakuin University, Faculty of General Education,
Yokohama, Japan}~$^{J}$

\end{minipage}\\
\makebox[3em]{$^{32}$}
\begin{minipage}[t]{14cm}
{\it Moscow Engineering Physics Institute, Moscow, Russia}~$^{O}$

\end{minipage}\\
\makebox[3em]{$^{33}$}
\begin{minipage}[t]{14cm}
{\it Moscow State University, Institute of Nuclear Physics,
Moscow, Russia}~$^{P}$

\end{minipage}\\
\makebox[3em]{$^{34}$}
\begin{minipage}[t]{14cm}
{\it Max-Planck-Institut f\"ur Physik, M\"unchen, Germany}

\end{minipage}\\
\makebox[3em]{$^{35}$}
\begin{minipage}[t]{14cm}
{\it NIKHEF and University of Amsterdam, Amsterdam, Netherlands}~$^{Q}$

\end{minipage}\\
\makebox[3em]{$^{36}$}
\begin{minipage}[t]{14cm}
{\it Physics Department, Ohio State University,
Columbus, Ohio 43210, USA}~$^{A}$

\end{minipage}\\
\makebox[3em]{$^{37}$}
\begin{minipage}[t]{14cm}
{\it Department of Physics, University of Oxford,
Oxford, United Kingdom}~$^{D}$

\end{minipage}\\
\makebox[3em]{$^{38}$}
\begin{minipage}[t]{14cm}
{\it INFN Padova, Padova, Italy}~$^{B}$

\end{minipage}\\
\makebox[3em]{$^{39}$}
\begin{minipage}[t]{14cm}
{\it Dipartimento di Fisica dell' Universit\`a and INFN,
Padova, Italy}~$^{B}$

\end{minipage}\\
\makebox[3em]{$^{40}$}
\begin{minipage}[t]{14cm}
{\it Department of Physics, Pennsylvania State University, University Park,\\
Pennsylvania 16802, USA}~$^{F}$

\end{minipage}\\
\makebox[3em]{$^{41}$}
\begin{minipage}[t]{14cm}
{\it Polytechnic University, Sagamihara, Japan}~$^{J}$

\end{minipage}\\
\makebox[3em]{$^{42}$}
\begin{minipage}[t]{14cm}
{\it Dipartimento di Fisica, Universit\`a 'La Sapienza' and INFN,
Rome, Italy}~$^{B}$

\end{minipage}\\
\makebox[3em]{$^{43}$}
\begin{minipage}[t]{14cm}
{\it Rutherford Appleton Laboratory, Chilton, Didcot, Oxon,
United Kingdom}~$^{D}$

\end{minipage}\\
\makebox[3em]{$^{44}$}
\begin{minipage}[t]{14cm}
{\it Raymond and Beverly Sackler Faculty of Exact Sciences, School of Physics, \\
Tel Aviv University, Tel Aviv, Israel}~$^{R}$

\end{minipage}\\
\makebox[3em]{$^{45}$}
\begin{minipage}[t]{14cm}
{\it Department of Physics, Tokyo Institute of Technology,
Tokyo, Japan}~$^{J}$

\end{minipage}\\
\makebox[3em]{$^{46}$}
\begin{minipage}[t]{14cm}
{\it Department of Physics, University of Tokyo,
Tokyo, Japan}~$^{J}$

\end{minipage}\\
\makebox[3em]{$^{47}$}
\begin{minipage}[t]{14cm}
{\it Tokyo Metropolitan University, Department of Physics,
Tokyo, Japan}~$^{J}$

\end{minipage}\\
\makebox[3em]{$^{48}$}
\begin{minipage}[t]{14cm}
{\it Universit\`a di Torino and INFN, Torino, Italy}~$^{B}$

\end{minipage}\\
\makebox[3em]{$^{49}$}
\begin{minipage}[t]{14cm}
{\it Universit\`a del Piemonte Orientale, Novara, and INFN, Torino,
Italy}~$^{B}$

\end{minipage}\\
\makebox[3em]{$^{50}$}
\begin{minipage}[t]{14cm}
{\it Department of Physics, University of Toronto, Toronto, Ontario,
Canada M5S 1A7}~$^{N}$

\end{minipage}\\
\makebox[3em]{$^{51}$}
\begin{minipage}[t]{14cm}
{\it Physics and Astronomy Department, University College London,
London, United Kingdom}~$^{D}$

\end{minipage}\\
\makebox[3em]{$^{52}$}
\begin{minipage}[t]{14cm}
{\it Warsaw University, Institute of Experimental Physics,
Warsaw, Poland}

\end{minipage}\\
\makebox[3em]{$^{53}$}
\begin{minipage}[t]{14cm}
{\it Institute for Nuclear Studies, Warsaw, Poland}

\end{minipage}\\
\makebox[3em]{$^{54}$}
\begin{minipage}[t]{14cm}
{\it Department of Particle Physics, Weizmann Institute, Rehovot,
Israel}~$^{S}$

\end{minipage}\\
\makebox[3em]{$^{55}$}
\begin{minipage}[t]{14cm}
{\it Department of Physics, University of Wisconsin, Madison,
Wisconsin 53706, USA}~$^{A}$

\end{minipage}\\
\makebox[3em]{$^{56}$}
\begin{minipage}[t]{14cm}
{\it Department of Physics, York University, Ontario, Canada M3J
1P3}~$^{N}$

\end{minipage}\\
\vspace{30em} \pagebreak[4]


\makebox[3ex]{$^{ A}$}
\begin{minipage}[t]{14cm}
 supported by the US Department of Energy\
\end{minipage}\\
\makebox[3ex]{$^{ B}$}
\begin{minipage}[t]{14cm}
 supported by the Italian National Institute for Nuclear Physics (INFN) \
\end{minipage}\\
\makebox[3ex]{$^{ C}$}
\begin{minipage}[t]{14cm}
 supported by the German Federal Ministry for Education and Research (BMBF), under
 contract No. 05 H09PDF\
\end{minipage}\\
\makebox[3ex]{$^{ D}$}
\begin{minipage}[t]{14cm}
 supported by the Science and Technology Facilities Council, UK\
\end{minipage}\\
\makebox[3ex]{$^{ E}$}
\begin{minipage}[t]{14cm}
 supported by an FRGS grant from the Malaysian government\
\end{minipage}\\
\makebox[3ex]{$^{ F}$}
\begin{minipage}[t]{14cm}
 supported by the US National Science Foundation. Any opinion,
 findings and conclusions or recommendations expressed in this material
 are those of the authors and do not necessarily reflect the views of the
 National Science Foundation.\
\end{minipage}\\
\makebox[3ex]{$^{ G}$}
\begin{minipage}[t]{14cm}
 supported by the Polish Ministry of Science and Higher Education as a scientific project No.
 DPN/N188/DESY/2009\
\end{minipage}\\
\makebox[3ex]{$^{ H}$}
\begin{minipage}[t]{14cm}
 supported by the Polish Ministry of Science and Higher Education
 as a scientific project (2009-2010)\
\end{minipage}\\
\makebox[3ex]{$^{ I}$}
\begin{minipage}[t]{14cm}
 supported by the German Federal Ministry for Education and Research (BMBF), under
 contract No. 05h09GUF, and the SFB 676 of the Deutsche Forschungsgemeinschaft (DFG) \
\end{minipage}\\
\makebox[3ex]{$^{ J}$}
\begin{minipage}[t]{14cm}
 supported by the Japanese Ministry of Education, Culture, Sports, Science and Technology
 (MEXT) and its grants for Scientific Research\
\end{minipage}\\
\makebox[3ex]{$^{ K}$}
\begin{minipage}[t]{14cm}
 supported by the Korean Ministry of Education and Korea Science and Engineering
 Foundation\
\end{minipage}\\
\makebox[3ex]{$^{ L}$}
\begin{minipage}[t]{14cm}
 supported by FNRS and its associated funds (IISN and FRIA) and by an Inter-University
 Attraction Poles Programme subsidised by the Belgian Federal Science Policy Office\
\end{minipage}\\
\makebox[3ex]{$^{ M}$}
\begin{minipage}[t]{14cm}
 supported by the Spanish Ministry of Education and Science through funds provided by
 CICYT\
\end{minipage}\\
\makebox[3ex]{$^{ N}$}
\begin{minipage}[t]{14cm}
 supported by the Natural Sciences and Engineering Research Council of Canada (NSERC) \
\end{minipage}\\
\makebox[3ex]{$^{ O}$}
\begin{minipage}[t]{14cm}
 partially supported by the German Federal Ministry for Education and Research (BMBF)\
\end{minipage}\\
\makebox[3ex]{$^{ P}$}
\begin{minipage}[t]{14cm}
 supported by RF Presidential grant N 41-42.2010.2 for the Leading
 Scientific Schools and by the Russian Ministry of Education and Science through its
 grant for Scientific Research on High Energy Physics\
\end{minipage}\\
\makebox[3ex]{$^{ Q}$}
\begin{minipage}[t]{14cm}
 supported by the Netherlands Foundation for Research on Matter (FOM)\
\end{minipage}\\
\makebox[3ex]{$^{ R}$}
\begin{minipage}[t]{14cm}
 supported by the Israel Science Foundation\
\end{minipage}\\
\makebox[3ex]{$^{ S}$}
\begin{minipage}[t]{14cm}
 supported in part by the MINERVA Gesellschaft f\"ur Forschung GmbH, the Israel Science
 Foundation (grant No. 293/02-11.2) and the US-Israel Binational Science Foundation \
\end{minipage}\\
\vspace{30em} \pagebreak[4]


\makebox[3ex]{$^{ a}$}
\begin{minipage}[t]{14cm}
also affiliated with University College London,
 United Kingdom\
\end{minipage}\\
\makebox[3ex]{$^{ b}$}
\begin{minipage}[t]{14cm}
now at University of Salerno, Italy\
\end{minipage}\\
\makebox[3ex]{$^{ c}$}
\begin{minipage}[t]{14cm}
now at Queen Mary University of London, United Kingdom\
\end{minipage}\\
\makebox[3ex]{$^{ d}$}
\begin{minipage}[t]{14cm}
also working at Max Planck Institute, Munich, Germany\
\end{minipage}\\
\makebox[3ex]{$^{ e}$}
\begin{minipage}[t]{14cm}
also Senior Alexander von Humboldt Research Fellow at Hamburg University,
 Institute of Experimental Physics, Hamburg, Germany\
\end{minipage}\\
\makebox[3ex]{$^{ f}$}
\begin{minipage}[t]{14cm}
also at Cracow University of Technology, Faculty of Physics,
 Mathemathics and Applied Computer Science, Poland\
\end{minipage}\\
\makebox[3ex]{$^{ g}$}
\begin{minipage}[t]{14cm}
supported by the research grant No. 1 P03B 04529 (2005-2008)\
\end{minipage}\\
\makebox[3ex]{$^{ h}$}
\begin{minipage}[t]{14cm}
now at Rockefeller University, New York, NY
 10065, USA\
\end{minipage}\\
\makebox[3ex]{$^{ i}$}
\begin{minipage}[t]{14cm}
now at DESY group FS-CFEL-1\
\end{minipage}\\
\makebox[3ex]{$^{ j}$}
\begin{minipage}[t]{14cm}
now at Institute of High Energy Physics, Beijing,
 China\
\end{minipage}\\
\makebox[3ex]{$^{ k}$}
\begin{minipage}[t]{14cm}
now at DESY group FEB, Hamburg, Germany\
\end{minipage}\\
\makebox[3ex]{$^{ l}$}
\begin{minipage}[t]{14cm}
also at Moscow State University, Russia\
\end{minipage}\\
\makebox[3ex]{$^{ m}$}
\begin{minipage}[t]{14cm}
now at University of Liverpool, United Kingdom\
\end{minipage}\\
\makebox[3ex]{$^{ n}$}
\begin{minipage}[t]{14cm}
on leave of absence at CERN, Geneva, Switzerland\
\end{minipage}\\
\makebox[3ex]{$^{ o}$}
\begin{minipage}[t]{14cm}
now at CERN, Geneva, Switzerland\
\end{minipage}\\
\makebox[3ex]{$^{ p}$}
\begin{minipage}[t]{14cm}
now at Goldman Sachs, London, UK\
\end{minipage}\\
\makebox[3ex]{$^{ q}$}
\begin{minipage}[t]{14cm}
also at Institute of Theoretical and Experimental Physics, Moscow, Russia\
\end{minipage}\\
\makebox[3ex]{$^{ r}$}
\begin{minipage}[t]{14cm}
also at INP, Cracow, Poland\
\end{minipage}\\
\makebox[3ex]{$^{ s}$}
\begin{minipage}[t]{14cm}
also at FPACS, AGH-UST, Cracow, Poland\
\end{minipage}\\
\makebox[3ex]{$^{ t}$}
\begin{minipage}[t]{14cm}
partially supported by Warsaw University, Poland\
\end{minipage}\\
\makebox[3ex]{$^{ u}$}
\begin{minipage}[t]{14cm}
now at Istituto Nucleare di Fisica Nazionale (INFN), Pisa, Italy\
\end{minipage}\\
\makebox[3ex]{$^{ v}$}
\begin{minipage}[t]{14cm}
now at Haase Energie Technik AG, Neum\"unster, Germany\
\end{minipage}\\
\makebox[3ex]{$^{ w}$}
\begin{minipage}[t]{14cm}
now at Department of Physics, University of Bonn, Germany\
\end{minipage}\\
\makebox[3ex]{$^{ x}$}
\begin{minipage}[t]{14cm}
also affiliated with DESY, Germany\
\end{minipage}\\
\makebox[3ex]{$^{ y}$}
\begin{minipage}[t]{14cm}
now at Japan Synchrotron Radiation Research Institute (JASRI), Hyogo, Japan\
\end{minipage}\\
\makebox[3ex]{$^{ z}$}
\begin{minipage}[t]{14cm}
also at University of Tokyo, Japan\
\end{minipage}\\
\makebox[3ex]{$^{\dagger}$}
\begin{minipage}[t]{14cm}
 deceased \
\end{minipage}\\
\makebox[3ex]{$^{aa}$}
\begin{minipage}[t]{14cm}
now at Kobe University, Japan\
\end{minipage}\\
\makebox[3ex]{$^{ab}$}
\begin{minipage}[t]{14cm}
supported by DESY, Germany\
\end{minipage}\\
\makebox[3ex]{$^{ac}$}
\begin{minipage}[t]{14cm}
supported by the Bogolyubov Institute for Theoretical Physics of the National
 Academy of Sciences, Ukraine\
\end{minipage}\\
\makebox[3ex]{$^{ad}$}
\begin{minipage}[t]{14cm}
STFC Advanced Fellow\
\end{minipage}\\
\makebox[3ex]{$^{ae}$}
\begin{minipage}[t]{14cm}
nee Korcsak-Gorzo\
\end{minipage}\\
\makebox[3ex]{$^{af}$}
\begin{minipage}[t]{14cm}
This material was based on work supported by the
 National Science Foundation, while working at the Foundation.\
\end{minipage}\\
\makebox[3ex]{$^{ag}$}
\begin{minipage}[t]{14cm}
also at Max Planck Institute, Munich, Germany, Alexander von Humboldt
 Research Award\
\end{minipage}\\
\makebox[3ex]{$^{ah}$}
\begin{minipage}[t]{14cm}
now at Nihon Institute of Medical Science, Japan\
\end{minipage}\\
\makebox[3ex]{$^{ai}$}
\begin{minipage}[t]{14cm}
now at SunMelx Co. Ltd., Tokyo, Japan\
\end{minipage}\\
\makebox[3ex]{$^{aj}$}
\begin{minipage}[t]{14cm}
now at Osaka University, Osaka, Japan\
\end{minipage}\\
\makebox[3ex]{$^{ak}$}
\begin{minipage}[t]{14cm}
also at \L\'{o}d\'{z} University, Poland\
\end{minipage}\\
\makebox[3ex]{$^{al}$}
\begin{minipage}[t]{14cm}
member of \L\'{o}d\'{z} University, Poland\
\end{minipage}\\
\makebox[3ex]{$^{am}$}
\begin{minipage}[t]{14cm}
now at Lund University, Lund, Sweden\
\end{minipage}\\
\makebox[3ex]{$^{an}$}
\begin{minipage}[t]{14cm}
also at University of Podlasie, Siedlce, Poland\
\end{minipage}\\

}


%% file: DESY-10-129-txt.tex
\pagenumbering{arabic} 
\pagestyle{plain}

\section{Introduction}
\label{sec-intro}

Deep inelastic scattering (DIS) of leptons off nucleons has proved to
be a key process in the understanding of the structure of the proton
and the Standard Model (SM). Neutral current (NC) DIS is mediated by 
the exchange of photons and $Z$ bosons and is sensitive to all quark
flavours. In contrast, at leading order, only down-type quarks and up-type
antiquarks contribute to $e^+p$ charged current (CC) DIS.  Thus this
process is a powerful probe of flavour-specific  parton distribution
functions (PDFs). The SM predicts that the cross section for charged
current $ep$ DIS depends linearly on the longitudinal polarisation of 
the incoming lepton beam.  The cross section becomes zero for right-handed
(left-handed) electron (positron) beams, due to the chiral nature of
the weak interaction.

Using data taken at the HERA $ep$ collider in the years $1994\rnge2000$ 
and $2004\rnge2006$, the H1 and ZEUS
collaborations have reported measurements of the cross sections for
CC DIS~\mcite{pl:b324:241,zfp:c67:565,pl:b379:319,epj:c13:609,epj:c19:269,epj:c21:33,epj:c30:1,pl:b634:173,prl:75:1006,zfp:c72:47,epj:c12:411,pl:b539:197,epj:c32:1,pl:b637:210,epj:c61:223}. 
These measurements extend the kinematic region
covered by fixed-target proton-structure
measurements~\mcite{zfp:c25:29,zfp:c49:187,zfp:c53:51,zfp:c62:575}  to
higher values of negative four-momentum-transfer squared, $Q^{2}$.

This paper presents measurements of the cross sections for $e^{+}p$ CC
DIS with a longitudinally polarised positron beam.  The measured cross
sections are compared with the SM predictions and previous ZEUS
measurements of $e^{+}p$ CC DIS with an unpolarised
positron beam~\mcite{epj:c32:1}.  Similar results in $e^{-}p$ CC DIS
have been published by the ZEUS Collaboration~\cite{epj:c61:223}.
The total $e^{+}p$ cross section in bins of polarisation is fitted and extrapolated 
to find the cross section for a fully left-handed polarised positron beam.
The upper limit on this cross section is used to extract a lower
limit on the mass of a hypothetical $W$ boson which couples to right-handed 
particles.

This analysis is based on a data set with a five-fold increase in 
integrated luminosity compared to the previously published analysis 
of polarised  $e^{+}p$ CC DIS~\mcite{pl:b637:210} and twice the integrated 
luminosity compared to the previously most precise published analysis 
of  $e^{+}p$ CC DIS (with unpolarised positrons)~\mcite{epj:c32:1}.

\section{Kinematic variables and cross sections}
\label{sec-kin}

Inclusive deep inelastic lepton-proton scattering can be described in
terms of the kinematic variables $x$, $y$ and $Q^2$.  The variable
$Q^2$ is defined as $Q^2 = -q^2 = -(k-k')^2$ where $k$ and $k'$ are
the four-momenta of the incoming and scattered lepton,
respectively. Bjorken $x$ is defined as $x=Q^2/2P \cdot q$ where $P$
is the four-momentum of the incoming proton. The variable $y$ is
defined as $y=P\cdot q/P \cdot k$. The variables $x$, $y$ and $Q^2$
are related by $Q^2=sxy$, where $s=4E_e E_p$ is the square of the
lepton-proton centre-of-mass energy (neglecting the masses of the
incoming particles) and $E_{e}$ and $E_{p}$ are the energies of the
incoming positron and proton, respectively.

The electroweak Born-level cross section for the CC reaction,
$e^{+}p\rightarrow \overline{\nu}_{e}X$, with a longitudinally polarised positron
beam can be expressed as~\cite{devenish:2003:dis}
\begin{equation}
\frac{d^2 \sigma ^{\rm CC}}{dxdQ^2}=(1+P_{e})\frac{G_{F}^{2}}{4\pi
  x}\bigg(\frac{M_{W}^{2}}{M_{W}^{2}+Q^{2}}\bigg) ^{2} \bigg[
  Y_{+}F_{2}^{\rm CC}(x,Q^{2})-Y_{-}xF_{3}^{\rm CC}(x,Q^{2})
  -y^{2}F_{L}^{\rm CC}(x,Q^{2})
  \bigg], \nonumber
\end{equation}
where $G_{F}$ is the Fermi constant, $M_{W}$ is the mass of the $W$
boson and $Y_{\pm}=1\pm(1-y)^{2}$.   The
longitudinal polarisation of the positron beam, $P_{e}$, is defined as
\begin{equation}
P_{e}=\frac{N_{R}-N_{L}}{N_{R}+N_{L}}, \nonumber
\label{eqn:pol}
\end{equation}
where $N_{R}$ and $N_{L}$ are the numbers of right- and left-handed
positrons in the beam.

The longitudinal structure function, 
$F_{L}^{\rm CC}$, is negligible except at values of $y$ close to 1. 
At leading order in QCD, the structure functions
$F_{2}^{\rm CC}$ and $xF_{3}^{\rm CC}$  for $e^{+}p$ collisions may be
written in terms of sums and differences of  quark and anti-quark 
PDFs as follows:
\begin{equation}
F_{2}^{\rm CC} = x[d(x,Q^{2}) + s(x,Q^{2}) + \bar{u}(x,Q^{2}) +
  \bar{c}(x,Q^{2})], \nonumber
\end{equation}
\begin{equation}
xF_{3}^{\rm CC} = x[d(x,Q^{2}) + s(x,Q^{2}) - \bar{u}(x,Q^{2}) -
  \bar{c}(x,Q^{2})], \nonumber
\end{equation}
where, for example, the PDF $d(x,Q^{2})$ gives the number density of
down quarks with momentum-fraction $x$ at a given $Q^{2}$.  Since the
top-quark mass is large and the off-diagonal elements of the CKM
matrix are small~\cite{Amsler:2008zzb}, the contribution from
third-generation quarks may be ignored~\cite{katz:2000:hera}.

The reduced cross section, $\tilde{\sigma}$, is
defined as
\begin{equation}
   \tilde{\sigma} = \left[\frac{G^2_F}{2 \pi x } \Biggl(
     \frac{M^2_W}{M^2_W + Q^2}\Biggr)^2 \right]^{-1}
         {\frac{d^2\sigma ^{\rm CC}}{dx \, dQ^2}}. \nonumber
\end{equation}

At leading order in QCD, the unpolarised reduced cross section 
depends on the quark momentum distributions as follows:
\begin{equation}
  \tilde{\sigma} (e^+ p \rightarrow {\overline{\nu}}_e X) = x\left[(\bar{u} + \bar{c} +
    (1-y)^2 (d + s) \right]. \nonumber
\end{equation}
\section{Experimental apparatus}
\label{sec-ccdet}

A detailed description of the ZEUS detector can be found
elsewhere~\cite{zeus:1993:bluebook}.  A brief outline of the
components most relevant for this analysis  is given below.

In the kinematic range of the analysis, charged particles were tracked
in the central tracking detector (CTD)~\citeCTD, the microvertex
detector (MVD)~\citeMVD and the straw tube tracker (STT)~\cite{nim:a535:191}. 
The CTD and 
the MVD operated in a magnetic field of $1.43\Tesla$ provided by a
thin superconducting solenoid. The CTD consisted of
72~cylindrical drift chamber  layers, organised in nine~superlayers
covering the  polar-angle\ZcoosysfnB~region
\mbox{$15^\circ<\theta<164^\circ$}. The MVD silicon tracker consisted
of a barrel (BMVD) and a forward (FMVD) section. The BMVD provided
polar-angle coverage for tracks with three measurements from
$30^\circ$ to $150^\circ$. The FMVD extended the polar-angle coverage
in the forward region down to $7^\circ$. The STT consisted of 48 sectors of
 two different sizes. Each sector contained 192 (small sector) or 264 
(large sector) straws of diameter
7.5 mm arranged into 3 layers. The sectors were trapezoidal in shape
and each subtended an azimuthal angle of $60^{\circ}$; six sectors
formed a superlayer. A particle passing through the complete
STT traversed 8 superlayers, which were rotated around the beam
direction at angles of 30$^{\circ}$ or 15$^{\circ}$ to each other. The STT
covered the polar-angle region $5^{\circ} < \theta<  23^{\circ}$.

The high-resolution uranium--scintillator calorimeter (CAL)~\citeCAL
consisted  of three parts: the forward (FCAL), the barrel (BCAL) and
the rear (RCAL) calorimeter, covering 99.7\% of the solid angle around
the nominal interaction point.  Each part was subdivided transversely
into towers and longitudinally into one electromagnetic section (EMC)
and either one (in RCAL) or two (in BCAL and FCAL) hadronic sections
(HAC). The smallest subdivision of the calorimeter was called a cell.
The CAL relative energy resolutions,  as measured under test-beam
conditions, were $\sigma(E)/E=0.18/\sqrt{E}$ for positrons and
$\sigma(E)/E=0.35/\sqrt{E}$ for hadrons, with $E$ in $\Gev$. The
timing resolution of the CAL was better than 1~ns  for energy deposits
exceeding 4.5~\gev.

An iron structure that surrounded the CAL was instrumented as a
backing calorimeter (BAC)~\cite{nim:a313:126} to measure energy
leakage from the CAL. Muon chambers in the forward~\cite{zeus:1993:bluebook}, 
barrel and rear regions~\cite{nim:a333:342} were used in this analysis to veto
background events induced by  cosmic-ray or beam-halo muons.

The luminosity was measured using the Bethe-Heitler reaction $ep
\rightarrow e \gamma p$ with the luminosity detector which consisted
of two independent systems, a photon calorimeter~\cite{desy-92-066,zfp:c63:391,acpp:b32:2025} and a magnetic spectrometer~\cite{Helbich:2005qf}.

The lepton beam in HERA became naturally transversely polarised
through the Sokolov-Ternov effect~\cite{sovpdo:8:1203,sovjnp:b9:238}.
The characteristic build-up time for the HERA accelerator was
approximately 40~minutes.  Spin rotators on either side of the ZEUS
detector changed the  transverse polarisation of the beam into
longitudinal polarisation and back again. The positron beam
polarisation was measured using  two independent polarimeters, the
transverse polarimeter (TPOL)~\cite{nim:a329:79} and the longitudinal
polarimeter (LPOL)~\cite{nim:a479:334}.  Both devices exploited the
spin-dependent cross section for Compton scattering of circularly
polarised photons off positrons to measure the beam polarisation.  The
luminosity and polarisation measurements were made over time intervals that
were much shorter than the polarisation build-up time.

The measurements are based on data samples collected with the ZEUS
detector in 2006 and 2007 when HERA collided protons of energy
$920\gev$ with positrons of energy $27.5\gev$, yielding collisions at
a centre-of-mass energy of $318\gev$. The integrated luminosities of
the data sample were $75.8 \pb^{-1}$ and $56.0 \pb^{-1}$ at mean
luminosity-weighted polarisations of $+$0.33 and $-$0.36,
respectively. Runs with mean absolute polarisation less than 15\% were 
rejected so that the polarisation measurement was reliable with a well 
understood systematic uncertainty. Figure~\ref{fig-lumipol} shows the luminosity
collected as a function of the  longitudinal polarisation of the
positron beam. 

\section{Monte Carlo simulation}
\label{sec-mc}

Monte Carlo (MC) simulation was used to determine the efficiency for
selecting events, the accuracy of kinematic  reconstruction, to
estimate the background rate and to extract cross sections for the full
kinematic region from the data.  A sufficient number of events was
generated to ensure that uncertainties from MC  statistics were negligible.
The MC samples were normalised to the total integrated luminosity of
the data.

Charged current DIS events, including electroweak radiative effects,
were  simulated using the {\sc Heracles}
4.6.6~\cite{cpc:69:155,*spi:www:heracles}  program with the {\sc
  Djangoh} 1.6~\cite{spi:www:djangoh11} interface to the MC generators
that provide the hadronisation. Initial-state radiation, vertex and
propagator corrections and two-boson exchange are included in {\sc
  Heracles}.  The parameters of the SM were set to the
PDG~\cite{Amsler:2008zzb} values.  The events were generated using the
CTEQ5D~\cite{epj:c12:375} PDFs. The colour-dipole model of {\sc
  Ariadne} 4.12~\cite{cpc:71:15} was used to simulate
$\mathcal{O}(\alpha_{S})$ plus leading-logarithmic corrections to the
result of the quark-parton model.  {\sc Ariadne} uses the Lund string model 
of {\sc Jetset} 7.4.1~\cite{cpc:39:347,*cpc:43:367,*cpc:82:74} for the 
hadronisation. A set of NC DIS events generated with {\sc Djangoh} was 
used to estimate the NC contamination in the CC sample. Photoproduction 
background was estimated using events simulated with {\sc Herwig}
5.9~\cite{cpc:67:465}.  Events simulated with {\sc Grape}
1.1~\cite{cpc:136:126} and {\sc Epvec} 1.0~\cite{np:b375:3}  were used
to estimate the background contribution from  di-lepton and single-$W$
production, respectively.

The ZEUS detector response was simulated using a program based on {\sc
  Geant} 3.21~\cite{tech:cern-dd-ee-84-1}.  The generated events were
passed through the detector simulation, subjected to the same trigger
requirements as the data and processed by the same reconstruction
programs.

\section{Reconstruction of kinematic variables}
\label{sec-ccrecon}

The main experimental signature of CC DIS events at HERA is 
large missing transverse momentum, $\PTMvector$. Figure \ref{fig-event}
shows such an event as  observed using the ZEUS detector.  The
struck quark gives rise to one or more jets of hadrons and the
energetic final-state neutrino escapes detection, leaving a large
imbalance in the transverse momentum observed in the detector. The
vector $\PTMvector$ is derived from the total visible hadronic momentum vector, 
$\PTvector$, by $\PTMvector  = -\PTvector$, where
\begin{equation}
\PTvector = \left(P_x, P_y\right) = 
  \left(\sum\limits_{i} E_i \sin \theta_i \cos \phi_i\;,\;
  \sum\limits_{i} E_i \sin \theta_i \sin \phi_i \right).\nonumber
  \label{eq:pt}
\end{equation}
The sums run over all CAL energy deposits, $E_i$, and $\theta_i$ and
$\phi_i$ are the polar and azimuthal angles of the
calorimeter deposit $i$ as viewed from the interaction vertex~\cite{epj:c61:223}. The polar angle of the hadronic system, $\gamma_h$, is defined as
\begin{equation}
\cos\gamma_h = \frac{\left(\overrightarrow{P}_{T}\right)^2 - \delta^2}
{\left(\overrightarrow{P}_{T}\right)^2 + \delta^2}, \nonumber
\end{equation}
where $\delta = \sum\limits_{i} E_i ( 1 - \cos \theta_{i} )  =
\sum\limits_{i} (E-P_Z)_{i}$.  In the naive quark-parton model,
$\gamma_h$ is the angle of the scattered quark.  Finally, the total
transverse energy, $E_T$, is given by $E_T    = \sum\limits_{i} E_i
\sin \theta_i$. 

The ratio of the parallel, $V_{P}$, and antiparallel, $V_{AP}$,
components of the hadronic transverse momentum can be used to
distinguish CC DIS from photoproduction events.  These variables are
defined as

\begin{equation}
V_{P} = \sum\limits_{i} \overrightarrow{P}_{T,i} \cdot
\overrightarrow{n}~~~{\rm for}~~\overrightarrow{P}_{T,i}
\cdot \overrightarrow{n}>0, \nonumber
\end{equation}
\begin{equation}
V_{AP} = - \sum\limits_{i} \overrightarrow{P}_{T,i} \cdot
\overrightarrow{n}~~~{\rm for}~~\overrightarrow{P}_{T,i}
\cdot \overrightarrow{n}<0, \nonumber
\end{equation}
where the sums are performed over all calorimeter deposits and
$\overrightarrow{n} =\overrightarrow{P}_{T}/P_{\;T}$.

 The kinematic variables were reconstructed using the Jacquet-Blondel
 method \cite{proc:epfacility:1979:391}: $y_{\rm{JB}} = \delta/(2E_e)$, 
$Q^2_{\rm{JB}} = \PTM^2/(1-y_{\rm{JB}})$, and $x_{\rm{JB}} = Q^2_{\rm{JB}}/(sy_{\rm{JB}})$. The resolution in $Q^2$ is $\approx$ $24\%$. The resolution in $x$ improves 
from $\approx$ $26$\% at $x=0.0078$ to $\approx$ $9\%$ at $x=0.65$.
The resolution in $y$ ranges from $\approx$ $15\%$ at $y=0.05$ to $\approx$ 
$8\%$ at $y=0.83$.
\section{Charged current event selection}
\label{sec-evsel}

Charged current DIS candidate events were selected by requiring a large
$\PTM$ in the event.  Backgrounds to CC DIS arise from  high-$E_T$ 
events in which the finite energy
resolution of the CAL or energy that escapes detection can lead to
significant missing transverse momentum.  Non-$ep$ events such
as beam-gas interactions, beam-halo muons or cosmic rays can also
cause substantial  imbalance in the measured transverse momentum and
constitute additional sources of background.  The following criteria
were imposed to select CC DIS events and reject these backgrounds.

\subsection{Trigger selection}
\label{subsec-Trigger}
Events were selected using the ZEUS three-level trigger
system~\cite{zeus:1993:bluebook,uproc:chep:1992:222,nim:a580:1257}.
At the first level, coarse calorimeter and tracking information
was available.  Events were selected using criteria based on the
energy, transverse  energy and missing transverse momentum measured in
the calorimeter.  Generally, events were triggered with low thresholds
on these quantities if a coincidence with CTD tracks from the event
vertex occurred, while higher  thresholds were required for events
with no CTD  tracks.

At the second level, timing information from the calorimeter was used
to reject events inconsistent with the bunch-crossing time.  In
addition, the topology of the CAL energy deposits was used to reject
background events.  In particular, a tighter cut was made on missing
transverse momentum, since the resolution in this variable was better
at the second than at the first level.

At the third level, full track reconstruction and vertex finding were
performed and used to reject candidate events with a vertex
inconsistent with an $ep$ interaction. Cuts were applied to
calorimeter quantities and reconstructed tracks to reduce
beam-gas contamination further.
 
\subsection{Offline selection}
For all events,  the
kinematic variables were recalculated using the $Z$-coordinate of the
event vertex ($Z_{\rm vtx}$) determined from charged-particle tracks. 
The requirements for event selection are given below:

\begin{itemize}
\item {kinematic cuts:  events were required to satisfy
  $Q^{2}_{\rm JB}>200 \gev^2$ and $y_{\rm JB}<0.9$.
  These requirements
  restricted the event sample to a region where the
  resolution of the kinematic quantities is good and the background
  is small;}
\item {missing transverse momentum: $\PTM>12 \gev$ was required and,
  in addition, the missing  transverse momentum excluding the
  calorimeter cells adjacent to the forward beam hole,  $\PTM'$, was
  required to exceed $10 \gev$;}
\item {primary interaction vertex: events were required
    to satisfy $| Z_{\rm vtx} | < 30$~cm. The improved tracking information 
    compared to the previous charged current analysis \cite{epj:c61:223} 
    allowed the 
    requirement of a reconstructed primary vertex in the full phase-space.
    This requirement strongly suppressed non-$ep$ backgrounds;}
  \item {rejection of photoproduction and di-leptons:  for events with
    $\PTM<20 \gev$, $V_{AP}/V_{P}$ $<0.25$ was required; for all other
    events, $V_{AP}/V_{P}<0.35$ was required. These requirements
    demanded an azimuthally collimated energy flow.  In addition, for
    all events, the azimuthal-angle difference, $\Delta\phi$, between the missing
    transverse  momentum measured by the tracks and that measured by
    the calorimeter was required to be less than $90^\circ$ for all events;}
  \item {rejection of NC DIS: NC DIS events with a poorly measured
    scattered positron or hadronic  jet can have significant  missing
    transverse momentum. Events with  $\delta > 30 \gev$ and an
    isolated electromagnetic  cluster in the calorimeter~\cite{nim:a365:508,nim:a391:360} 
    were rejected as detailed in a previous publication~\cite{epj:c61:223};}
  \item {rejection of remaining non-$ep$ background: interactions between the
    beams and residual gas in the beam pipe  or upstream accelerator
    components can lead to events with significant missing transverse
    momentum. However, for these interactions, the arrival times of
    energy deposits in the  calorimeter are inconsistent with the
    bunch-crossing time and were used to reject such
    events.  Events caused by interactions with the residual gas are
    characterised by a large fraction of tracks not associated with
    the $ep$ interaction vertex; such events were rejected by
    applying a cut in two dimensions on the number of vertex tracks, 
    $N_{\rm VtxTrks}$, versus the total number of tracks, $N_{\rm Trks}$.
    This cut was $N_{\rm VtxTrks}>0.125\cdot (N_{\rm Trks}-20)$. Vertex tracks
    were required to originate in the MVD or in the first superlayer of the CTD and
    to have a polar angle in the range of $ 15^\circ < \theta < 160^\circ $.
    Requirements on energy fractions in the calorimeter cells plus
    muon-finding algorithms based on tracking, calorimeter and muon
    chamber  information were used to reject events caused by cosmic
    rays or muons in the beam halo.}
\end{itemize}

A total of 2327 data events satisfied all criteria in the
positive-polarisation sample and 821 events in the
negative-polarisation sample.  The background contamination was
estimated to be typically less than 1.5\%,  but reached 8\% in
the lowest-$Q^{2}$ bin and 21\% in the lowest-$x$ bin of the 
negative-polarisation sample.
Similarly, it was typically less than 1\% but reached almost 4\% in
the lowest-$Q^{2}$ bin and 10\% in the lowest-$x$ bin of the positive-polarisation
sample. For the combined sample (positive and negative polarisations) 
the estimated number of background events 
was 19, 11 and 6.6 for photoproduction, single-$W$ production and di-lepton 
events, respectively. The di-lepton background was dominated by $\mu\mu$ and 
$\tau\tau$ events. The contamination from NC events was estimated to be very 
small (0.7 events for the combined sample). Non-$ep$ backgrounds were 
negligible. Figure~\ref{fig-cc_ctrl} compares the distributions of data
events entering the final CC sample with the MC expectation  for the
sum of the CC signal and $ep$ background events. The MC simulations
give a reasonable  description of the data.

\section{Cross-section determination}

The measured cross section in a particular kinematic bin, for example
in $d\sigma/dQ^{2}$, was determined from

\begin{equation}
\frac{d\sigma_{\rm Born}}{dQ^{2}}=\frac{N_{\rm data}-N_{\rm
    bg}}{N_{\rm MC}}\cdot\frac{d\sigma_{\rm Born}^{\rm SM}}{dQ^{2}},
\nonumber
\end{equation}

where $N_{\rm data}$ is the number of data events, $N_{\rm bg}$ is the
number of background events estimated from the MC simulation and
$N_{\rm MC}$ is the number of signal MC events. The Standard Model prediction,
$\frac{d\sigma_{\rm
    Born}^{\rm SM}}{dQ^{2}}$, is 
evaluated in the on-shell scheme using the PDG values for the
electroweak parameters and the same PDF set (CTEQ5D)~\cite{epj:c12:375} used to generate 
the MC data. A similar procedure was used for  
$d\sigma/dx$, $d\sigma/dy$ and the reduced cross 
section.  Consequently, the acceptance, as well as the
bin-centring and  radiative corrections were all taken from the MC
simulation.  The equation above includes the extrapolation of the 
single-differential cross-sections $ d \sigma/ d  Q^2$ and $ d \sigma/ d
x$ to the full $y$ range.

\section{Systematic uncertainties}
\label{sec-sys}
Different systematic uncertainties in the measured cross sections were
determined using one of two methods \cite{oliver:phd:2010}. 
The first set of systematic uncertainties relies on MC simulations 
and was calculated by changing relevant parameters of the analysis 
by their estimated errors  and repeating the extraction of the cross sections. 
The difference between the nominal cross section and that obtained from 
the modified analysis gave an estimate of the systematic uncertainty in 
each bin. The second method of calculating systematic uncertainties exploited
the similarity between NC and CC hadronic final states. 
The following systematics were determined using the first method: 
\begin{itemize}

\item{calorimeter energy scale: the relative uncertainty of the
  hadronic energy scale was 2\%.  
  The variation of the energy scale for each  of the
  calorimeters simultaneously up or down by this amount gave the
  systematic uncertainty  on the total measured energy in the
  calorimeter.  The resulting uncertainties in the measured cross
  sections were $\approx$ 1\% for the total cross sections 
  and for the single-differential cross sections were typically within  
 $\pm 3\%$, but increased to $\pm (25 - 33)\%$ in the highest-$Q^{2}$ and 
 highest-$x$ bins. The uncertainties reached $35\%$ in the highest-$Q^2$ and 
  highest-$x$ reduced cross-section bin;}
\item{efficiency of the FLT tracking: the charged current MC was 
corrected for observed differences in the CTD tracking efficiency 
between data and MC at the first-level trigger \cite{oliver:phd:2010}. 
The correction was derived from independent 
samples of NC data and NC MC events with the scattered electron removed 
in order to simulate CC events (pseudo-CC). The mean correction was 
$\approx$ 3.5\% for the positive-polarisation sample and $\approx$ 5\% for the 
negative-polarisation sample. The uncertainty on this correction was
  $50\%$ of its value. 
 The resulting uncertainties on the total cross sections were less
 than $1.5\%$ and for the single-differential and reduced cross sections 
 were typically $1-2 \%$ and were always less than $4\%$;}
\item{background subtraction: the uncertainty in the small
  contribution from  photoproduction was estimated. The 
  $V_{AP}/V_{P}$ distribution was plotted for data and MC events with all 
  selection cuts applied except for the cut on $V_{AP}/V_{P}$. A 
  $\chi^2$ fit of the MC to the data distribution was performed, varying the 
  normalisation of the photoproduction MC until it produced the best 
  description of the data. The fit resulted in a normalisation factor of 
  $0.880^{+0.090}_{-0.085}$.  The nominal photoproduction sample 
  was therefore scaled by a factor of 0.970 and by a factor of 0.795, 
  resulting in very small modifications of less than 0.2\% to the cross sections.} 
\end{itemize}

In the second method, a set of NC 
DIS data events with the scattered positron removed (pseudo-CC data) 
was reweighted to the $Q^2$ and $x$ of the CC DIS MC. 
In order to estimate the bias introduced into 
the  measurements from an imperfect description of the data by 
the MC simulation, the efficiencies  for each of the selection 
criteria were measured using the hadronic final state in NC DIS data 
and compared to those obtained with the CC MC. The differences in 
the efficiencies between the two samples were taken as estimates 
of the systematic uncertainties which were typically within $\pm 3\%$.

The individual uncertainties were added in quadrature separately for
the positive and negative deviations from the nominal cross section
values to obtain the total systematic uncertainty. 

The uncertainties on the electroweak corrections to CC DIS are less than 0.5\% 
\cite{Makarenko}. No uncertainty was included in the measured
cross sections from this source. 

The relative uncertainty in the measured polarisation was $3.6\%$
using the LPOL and $4.2\%$ using the TPOL. The choice of polarimeter
measurement was made on a run-by-run basis depending on which was
active the longer, in order to maximise the luminosity. For the final 
selection, the TPOL was used for 64\% (24\%) of the negative 
(positive) polarisation run period. The combined, luminosity-weighted 
systematic uncertainty on the polarisation measurement was 4.0\% (3.7\%) for
negative (positive) polarisation.  The uncertainty of $2.6\%$ on the
measured total luminosity was not included in the differential
cross-section figures or the tables.  
\section{Results}
\label{sec-res}

The total cross section, corrected to the Born level in the
electroweak interaction, for $e^+ p$ CC DIS in the kinematic region
$Q^{2}>200 \gev^{2}$ was measured to be
\begin{equation} 
\sigma^{\rm CC}(P_{e}=-0.36)=22.9\pm 0.82({\rm stat.}) \pm 0.60({\rm lumi.}) \pm 0.40({\rm syst.})~{\rm pb}, \nonumber
\end{equation} 
\begin{equation} 
\sigma^{\rm CC}(P_{e}=+0.33)=48.0\pm 1.01({\rm stat.}) \pm 1.25({\rm lumi.}) \pm 0.77({\rm syst.})~{\rm pb}. \nonumber
\end{equation}

The total cross section is shown as a
function of the longitudinal polarisation  of the lepton beam in
Fig.~\ref{fig-cctotal}, including previous ZEUS measurements  from
both $e^{-}p$ and $e^{+}p$
data~\cite{pl:b539:197,epj:c32:1,epj:c61:223} and previous H1 measurements from 
 $e^{+}p$ data~\cite{pl:b634:173}. The H1 measurements were scaled to 
the kinematic region of this analysis.  The uncertainty in the
 measured luminosity is included in the systematic uncertainty in 
Fig.~\ref{fig-cctotal}.
The data are compared to the SM predictions evaluated at
next-to-leading order in QCD~\cite{dispred} using the HERAPDF1.0~\cite{:2009wt},
ZEUS-JETS~\cite{epj:c42:1}, CTEQ6.6~\cite{Nadolsky:2008zw} and 
MSTW2008~\cite{Martin:2009iq} PDFs, which describe the data well.

The single-differential cross-sections $d\sigma/dQ^{2}$, $d\sigma/dx$
and $d\sigma/dy$  for CC DIS are shown in
Figs.~\ref{fig-dsdq2},~\ref{fig-dsdx} and~\ref{fig-dsdy} 
for $Q^2>200 \gev^2$ and given in
Tables~\ref{tab-single-diff}, \ref{tab-uncorr_single_pos} and
\ref{tab-uncorr_single_neg}, respectively. 
The cross sections are well described by the SM evaluated using the HERAPDF1.0, 
ZEUS-JETS, CTEQ6.6 and MSTW2008 PDFs. The precision of the data is comparable 
to the uncertainties in the SM predictions; therefore these data have the 
potential to constrain the PDFs further.

The reduced cross-section $\tilde{\sigma}$ was measured in the kinematic range
\mbox{$200<Q^2<60\,000 \gev^2$} and \mbox{$0.006<x<0.562$} and  is
shown as a function of $x$ at fixed values of $Q^{2}$ in
Figs.~\ref{fig-double-separate} and~\ref{fig-double} and given in
 Tables~\ref{tab-double},
\ref{tab-uncorr_double_pos} and \ref{tab-uncorr_double_neg}.  The data
points are shown separately for positive and negative polarisation in 
Fig.~\ref{fig-double-separate} and are shown for the entire data set in 
Fig.~\ref{fig-double}, corrected to $P_{e}=0$ using the SM prediction from
 {\sc Hector} using CTEQ5D PDFs. The predictions of the SM evaluated using the 
HERAPDF1.0, ZEUS-JETS, 
CTEQ6.6 and MSTW2008 PDFs give a good description of the data.  
The contributions from the PDF combinations $(d + s)$ and $(\bar{u} + \bar{c})$, 
obtained  in the \MSbar scheme from the HERAPDF1.0 PDFs, are shown separately.

The SM $W$ boson couples only to left-handed fermions and right-handed
anti-fermions. Therefore, the angular  distribution of the scattered
quark in $e^{+}\bar{q}$ CC DIS will be flat in the positron-quark
centre-of-mass scattering angle, $\theta^{*}$,  while it will exhibit
a $(1+\cos\theta^{*})^{2}$ distribution in $e^{+}q$ scattering.
Since $(1-y)^2 \propto (1+\cos\theta^{*})^{2}$, the helicity structure
of CC interactions can be illustrated  by plotting the reduced
 cross section  versus $(1-y)^2$ in bins of $x$, see Section \ref{sec-kin}. 
The measurement is shown in Fig. 10 and is well described by the SM.
At leading order in QCD, the intercept of the prediction gives  the 
($\bar{u}+\bar{c}$) contribution, while the slope gives the ($d+s$)
contribution.

The CC $e^+p$ DIS  cross section becomes zero for fully left-handed positron 
beams, thus a non-zero cross section at $P_e = -1$ might point to the 
existence of a right-handed $W$ boson, $W_R$, and right-handed neutrinos, 
$\nu_R$~\cite{Senjanovic:1975rk, PhysRevD.11.2558}. The program {\sc Hector} 
was used to calculate the cross section for right-handed CC interactions in 
$e^+p$ DIS as a function of the mass of the $W_R$, $M_{W_{R}}$. It was assumed 
that the coupling strength and propagator dependence on the mass of the boson 
are the same as in SM CC interactions. The outgoing right-handed neutrinos 
were assumed to be light.
A linear function was fit to the total cross section in 8 bins of polarisation,
 including the previous ZEUS measurement of unpolarised $e^+p$ CC DIS, and 
extrapolated to $P_e = -1$. The fit and extrapolation to $P_e = -1$ is shown 
in Fig.~\ref{fig-polbins}. The cross sections measured in each bin are given in 
Table~\ref{tab-polbins}.
The upper limit on the cross section was converted to a lower limit on $M_{W_{R}}$:
\begin{equation}
\sigma^{\rm CC}(P_{e}=-1) < 2.9~{\rm pb} {\textnormal{ at 95\% CL,}} \nonumber
\end{equation}
\begin{equation}
M_{W_{R}} > 198~{\rm GeV} {\textnormal{ at 95\% CL.}} \nonumber
\end{equation}

The limit on $M_{W_R}$ set in this analysis is complementary to the 
limits obtained from direct searches \cite{Amsler:2008zzb,Alitti:1993pn,Abazov:2008vj,Aaltonen:2009qu,:2007bs}. In the direct searches, the $W$ boson is time-like, 
whereas the limit from this analysis is for a space-like $W$. 
\section{Summary}
\label{sec-sum}

The cross sections for charged current  deep inelastic scattering in
$e^{+}p$ collisions with  longitudinally polarised positron beams have
been measured.  The measurements are based on a data sample with an
integrated luminosity  of 132~\pb$^{-1}$ collected with the ZEUS
detector at HERA  at a centre-of-mass energy of 318\gev.  The total
cross section is given for positive and negative  values of the
longitudinal polarisation of the positron beam. In addition, the
single-differential cross-sections $d\sigma/d Q^2$, $d\sigma/d x$ and
$d\sigma/d y$ for  $Q^{2}>200\gev^2$ are measured. The reduced cross
section is presented in the kinematic range  $200<Q^2<60\,000 \gev^2$ and
$0.006<x<0.562$.  The measured cross sections are well
described by the predictions of  the Standard Model. Finally, 
a lower limit on the mass of a hypothetical right-handed $W$ boson 
is extracted from the upper limit of the cross section at $P_e = -1$. 
The limit obtained is $M_{W_{R}} > 198~{\rm GeV}$ at 95\% CL.
\section*{Acknowledgements}

We appreciate the contributions to the construction and maintenance of 
the ZEUS detector of many people who are not listed as authors.
The HERA machine group and the 
DESY computing staff are especially acknowledged for their success in 
providing excellent operation of the collider and the data-analysis  
environment.  We thank the DESY directorate for their strong support and 
encouragement.  

\vfill\eject

%% file: DESY-10-129-ref.tex
{
\def\bibname{\Large\bf References}
\def\refname{\Large\bf References}
\pagestyle{plain}
\ifzeusbst
  \bibliographystyle{./BiBTeX/bst/l4z_default}
\fi
\ifzdrftbst
  \bibliographystyle{./BiBTeX/bst/l4z_draft}
\fi
\ifzbstepj
  \bibliographystyle{./BiBTeX/bst/l4z_epj}
\fi
\ifzbstnp
  \bibliographystyle{./BiBTeX/bst/l4z_np}
\fi
\ifzbstpl
  \bibliographystyle{./BiBTeX/bst/l4z_pl}
\fi
{\raggedright
\bibliography{./BiBTeX/user/syn.bib,%
             ./BiBTeX/bib/l4z_articles.bib,%
             ./BiBTeX/bib/l4z_books.bib,%
             ./BiBTeX/bib/l4z_conferences.bib,%
             ./BiBTeX/bib/l4z_h1.bib,%
             ./BiBTeX/bib/l4z_misc.bib,%
             ./BiBTeX/bib/l4z_old.bib,%
             ./BiBTeX/bib/l4z_preprints.bib,%
             ./BiBTeX/bib/l4z_replaced.bib,%
             ./BiBTeX/bib/l4z_temporary.bib,%
             ./BiBTeX/bib/l4z_zeus.bib,%
             ./BiBTeX/user/extraReferencesForCC.bib}}
}
\vfill\eject

%% file: DESY-10-129-tab.tex
\begin{table}[p]
\begin{center}

\begin{tabular}{|c|c|c|c|} 
\hline 
$Q^2$ range ($\gev^2$) & $Q^2$ ($\gev^2$) & \multicolumn{2}{c|}{$d\sigma/dQ^{2}$ (pb/$\gev^2$)}\\ 
\hline 
& & $P_{e}=+0.33$ & $P_{e}=-0.36$ \\ 
\hline 
$200-400$ & $280$ & $(4.21 ^{+0.27}_{-0.25}$ $^{+0.17} _{-0.18}) \cdot 10^{-2}$ & $(2.25^{+0.23}_{-0.21}$ $^{+0.09}_{-0.10}) \cdot 10^{-2}$ \\ 
$400-711$ & $530$ & $(3.19 ^{+0.16}_{-0.15}$ $^{+0.10} _{-0.10}) \cdot 10^{-2}$ & $(1.25^{+0.12}_{-0.11}$ $^{+0.04}_{-0.04}) \cdot 10^{-2}$ \\ 
$711-1265$ & $950$ & $(1.69 ^{+0.08}_{-0.08}$ $^{+0.03} _{-0.04}) \cdot 10^{-2}$ & $(8.45^{+0.70}_{-0.65}$ $^{+0.17}_{-0.21}) \cdot 10^{-3}$ \\ 
$1265-2249$ & $1700$ & $(8.87 ^{+0.43}_{-0.41}$ $^{+0.11} _{-0.14}) \cdot 10^{-3}$ & $(4.18^{+0.36}_{-0.33}$ $^{+0.07}_{-0.06}) \cdot 10^{-3}$ \\ 
$2249-4000$ & $3000$ & $(3.91 ^{+0.21}_{-0.20}$ $^{+0.10} _{-0.10}) \cdot 10^{-3}$ & $(1.97^{+0.18}_{-0.17}$ $^{+0.06}_{-0.06}) \cdot 10^{-3}$ \\ 
$4000-7113$ & $5300$ & $(1.30 ^{+0.09}_{-0.09}$ $^{+0.07} _{-0.07}) \cdot 10^{-3}$ & $(6.81^{+0.82}_{-0.73}$ $^{+0.39}_{-0.38}) \cdot 10^{-4}$ \\ 
$7113-12469$ & $9500$ & $(2.67 ^{+0.31}_{-0.28}$ $^{+0.30} _{-0.24}) \cdot 10^{-4}$ & $(9.66^{+2.40}_{-1.96}$ $^{+1.10}_{-0.84}) \cdot 10^{-5}$ \\ 
$12469-22494$ & $17000$ & $(3.17 ^{+0.79}_{-0.64}$ $^{+0.61} _{-0.50}) \cdot 10^{-5}$ & $(1.80^{+0.77}_{-0.56}$ $^{+0.34}_{-0.28}) \cdot 10^{-5}$ \\ 
$22494-60000$ & $30000$ & $(1.46 ^{+1.42}_{-0.79}$ $^{+0.48} _{-0.40}) \cdot 10^{-6}$ & $(1.33^{+1.76}_{-0.86}$ $^{+0.44}_{-0.37}) \cdot 10^{-6}$ \\ 
\hline 
$x$ range & $x$ & \multicolumn{2}{c|}{$d\sigma/dx$ (pb)}\\ 
\hline 
& & $P_{e}=+0.33$ & $P_{e}=-0.36$ \\ 
\hline 
$0.006-0.010$ & $0.0078$ & $(6.39 ^{+1.07}_{-0.93}$ $^{+0.42} _{-0.70}) \cdot 10^{2}$ & $(3.64^{+0.98}_{-0.79}$ $^{+0.25}_{-0.36}) \cdot 10^{2}$ \\ 
$0.010-0.021$ & $0.015$ & $(6.81 ^{+0.43}_{-0.40}$ $^{+0.26} _{-0.32}) \cdot 10^{2}$ & $(3.32^{+0.36}_{-0.33}$ $^{+0.14}_{-0.15}) \cdot 10^{2}$ \\ 
$0.021-0.046$ & $0.032$ & $(4.62 ^{+0.19}_{-0.19}$ $^{+0.09} _{-0.09}) \cdot 10^{2}$ & $(1.98^{+0.15}_{-0.14}$ $^{+0.04}_{-0.04}) \cdot 10^{2}$ \\ 
$0.046-0.100$ & $0.068$ & $(2.19 ^{+0.09}_{-0.08}$ $^{+0.03} _{-0.03}) \cdot 10^{2}$ & $(1.07^{+0.07}_{-0.07}$ $^{+0.01}_{-0.02}) \cdot 10^{2}$ \\ 
$0.100-0.178$ & $0.130$ & $(8.86 ^{+0.47}_{-0.45}$ $^{+0.20} _{-0.19}) \cdot 10^{1}$ & $(4.87^{+0.42}_{-0.39}$ $^{+0.12}_{-0.11}) \cdot 10^{1}$ \\ 
$0.178-0.316$ & $0.240$ & $(3.30 ^{+0.23}_{-0.22}$ $^{+0.14} _{-0.14}) \cdot 10^{1}$ & $(1.49^{+0.19}_{-0.17}$ $^{+0.07}_{-0.07}) \cdot 10^{1}$ \\ 
$0.316-0.562$ & $0.420$ & $(7.75 ^{+1.03}_{-0.92}$ $^{+0.70} _{-0.66}) \cdot 10^{0}$ & $(2.83^{+0.81}_{-0.64}$ $^{+0.27}_{-0.23}) \cdot 10^{0}$ \\ 
$0.562-1.000$ & $0.650$ & $(1.71 ^{+3.94}_{-1.42}$ $^{+0.58} _{-0.36}) \cdot 10^{-1}$ & $(2.35^{+5.41}_{-1.95}$ $^{+0.58}_{-0.51}) \cdot 10^{-1}$ \\ 
\hline 
$y$ range & $y$ & \multicolumn{2}{c|}{$d\sigma/dy$ (pb)}\\ 
\hline 
& & $P_{e}=+0.33$ & $P_{e}=-0.36$ \\ 
\hline 
$0.00-0.10$ & $0.05$ & $103.9^{+5.4} _{-5.1}$ $^{+1.5}_{-1.9}$ & $56.2^{+4.7} _{-4.4}$ $^{+1.0} _{-1.1}$ \\ 
$0.10-0.20$ & $0.15$ & $87.0^{+3.9} _{-3.7}$ $^{+0.9}_{-1.1}$ & $39.6^{+3.1} _{-2.9}$ $^{+0.6} _{-0.6}$ \\ 
$0.20-0.34$ & $0.27$ & $66.5^{+2.9} _{-2.8}$ $^{+0.9}_{-1.0}$ & $31.9^{+2.4} _{-2.3}$ $^{+0.5} _{-0.5}$ \\ 
$0.34-0.48$ & $0.41$ & $49.3^{+2.7} _{-2.6}$ $^{+0.9}_{-0.9}$ & $20.5^{+2.1} _{-1.9}$ $^{+0.5} _{-0.4}$ \\ 
$0.48-0.62$ & $0.55$ & $35.6^{+2.5} _{-2.3}$ $^{+0.9}_{-1.1}$ & $18.5^{+2.2} _{-1.9}$ $^{+0.5} _{-0.6}$ \\ 
$0.62-0.76$ & $0.69$ & $25.9^{+2.4} _{-2.2}$ $^{+1.1}_{-1.1}$ & $11.1^{+1.9} _{-1.7}$ $^{+0.5} _{-0.5}$ \\ 
$0.76-0.90$ & $0.83$ & $19.5^{+2.6} _{-2.3}$ $^{+1.5}_{-1.5}$ & $10.5^{+2.4} _{-2.0}$ $^{+0.9} _{-0.9}$ \\ 
\hline 
\end{tabular}

\caption{Values of the differential cross-sections $d \sigma /dQ^{2}$,  $d \sigma /dx$ and $d \sigma /dy$ for $P_{e}=+0.33 \pm 0.01$ and $P_{e}= -0.36 \pm 0.01$. The following quantities are given: the range of the measurement; the value at which the cross section is quoted and the measured cross section, with statistical and systematic uncertainties.}
\label{tab-single-diff}
\end{center}
\end{table}

\begin{table}[p]
\begin{center}

\begin{tabular}{|c|c|c|c|c|c|c|} 
\hline 
\multicolumn{7}{|c|}{$d\sigma/dQ^{2}$ ($P_{e}=+0.33 \pm 0.01$)}\\ 
\hline 
$Q^2$ ($\gev^2$)     & $d\sigma/dQ^{2}$ (pb/$\gev^2$)& $\delta_{{\rm stat}}$ (\%)&  $\delta_{{\rm syst}}$ (\%)&   $\delta_{{\rm unc}}$ (\%)& $\delta_{\rm trk}$ (\%)& $\delta_{\rm es}$(\%)\\ 
\hline 
 $ 280$ & $4.21\cdot 10^{-2}$ & $ ^{+ 6.4} _{-6.0}$ & $ ^{+4.0}_{-4.4}$ & $^{+0.5}_{-2.0}$ &$^{+1.4}_{-1.4}$ &$^{+3.7}_{-3.6}$ \\
 $ 530$ & $3.19\cdot 10^{-2}$ & $ ^{+ 5.1} _{-4.8}$ & $ ^{+3.1}_{-3.0}$ & $^{+0.6}_{-1.2}$ &$^{+1.3}_{-1.3}$ &$^{+2.7}_{-2.4}$ \\
 $ 950$ & $1.69\cdot 10^{-2}$ & $ ^{+ 4.9} _{-4.7}$ & $ ^{+1.9}_{-2.4}$ & $^{+0.6}_{-1.6}$ &$^{+1.2}_{-1.1}$ &$^{+1.3}_{-1.3}$ \\
 $ 1700$ & $8.87\cdot 10^{-3}$ & $ ^{+ 4.9} _{-4.7}$ & $ ^{+1.3}_{-1.6}$ & $^{+0.6}_{-1.2}$ &$^{+1.1}_{-1.1}$ &$^{+0.3}_{-0.0}$ \\
 $ 3000$ & $3.91\cdot 10^{-3}$ & $ ^{+ 5.5} _{-5.2}$ & $ ^{+2.5}_{-2.7}$ & $^{+0.6}_{-1.0}$ &$^{+1.0}_{-1.0}$ &$^{-2.2}_{+2.3}$ \\
 $ 5300$ & $1.30\cdot 10^{-3}$ & $ ^{+ 7.1} _{-6.7}$ & $ ^{+5.6}_{-5.2}$ & $^{+0.8}_{-0.6}$ &$^{+1.0}_{-0.9}$ &$^{-5.1}_{+5.5}$ \\
 $ 9500$ & $2.67\cdot 10^{-4}$ & $ ^{+ 11.7} _{-10.5}$ & $ ^{+11.3}_{-8.8}$ & $^{+0.9}_{-1.9}$ &$^{+0.9}_{-0.9}$ &$^{-8.5}_{+11.2}$ \\
 $ 17000$ & $3.17\cdot 10^{-5}$ & $ ^{+ 24.9} _{-20.3}$ & $ ^{+19.3}_{-15.9}$ & $^{+0.0}_{-4.7}$ &$^{+0.9}_{-0.9}$ &$^{-15.1}_{+19.3}$ \\
 $ 30000$ & $1.46\cdot 10^{-6}$ & $ ^{+ 97.3} _{-54.4}$ & $ ^{+32.6}_{-27.4}$ & $^{+0.0}_{-5.6}$ &$^{+1.0}_{-1.0}$ &$^{-26.8}_{+32.6}$ \\
\hline 
\multicolumn{7}{|c|}{$d\sigma/dx$ ($P_{e}=+0.33 \pm 0.01$)}\\ 
\hline 
$x$ & $d\sigma/dx$ (pb)& $\delta_{{\rm stat}}$ (\%)&  $\delta_{{\rm syst}}$ (\%)&   $\delta_{{\rm unc}}$ (\%)& $\delta_{\rm trk}$ (\%)& $\delta_{\rm es}$ (\%)\\ 
\hline 
 $ 0.0078$ & $6.39\cdot 10^{2}$ & $ ^{+ 16.8} _{-14.5}$ & $ ^{+6.5}_{-10.9}$ & $^{+0.7}_{-9.2}$ &$^{+2.4}_{-2.3}$ &$^{+5.9}_{-5.4}$ \\
 $ 0.015$ & $6.81\cdot 10^{2}$ & $ ^{+ 6.3} _{-5.9}$ & $ ^{+3.8}_{-4.7}$ & $^{+0.6}_{-3.4}$ &$^{+2.0}_{-1.9}$ &$^{+3.2}_{-2.7}$ \\
 $ 0.032$ & $4.62\cdot 10^{2}$ & $ ^{+ 4.2} _{-4.0}$ & $ ^{+1.9}_{-2.0}$ & $^{+0.6}_{-0.8}$ &$^{+1.4}_{-1.4}$ &$^{+1.0}_{-1.2}$ \\
 $ 0.068$ & $2.19\cdot 10^{2}$ & $ ^{+ 3.9} _{-3.8}$ & $ ^{+1.2}_{-1.2}$ & $^{+0.5}_{-0.6}$ &$^{+1.0}_{-1.0}$ &$^{-0.3}_{+0.4}$ \\
 $ 0.130$ & $8.86\cdot 10^{1}$ & $ ^{+ 5.3} _{-5.1}$ & $ ^{+2.2}_{-2.1}$ & $^{+0.6}_{-0.9}$ &$^{+0.7}_{-0.7}$ &$^{-1.8}_{+2.0}$ \\
 $ 0.240$ & $3.30\cdot 10^{1}$ & $ ^{+ 7.1} _{-6.7}$ & $ ^{+4.2}_{-4.3}$ & $^{+0.4}_{-1.3}$ &$^{+0.5}_{-0.5}$ &$^{-4.1}_{+4.1}$ \\
 $ 0.420$ & $7.75\cdot 10^{0}$ & $ ^{+ 13.3} _{-11.8}$ & $ ^{+9.1}_{-8.5}$ & $^{+0.7}_{-2.7}$ &$^{+0.4}_{-0.4}$ &$^{-8.1}_{+9.0}$ \\
 $ 0.650$ & $1.71\cdot 10^{-1}$ & $ ^{+ 229.9} _{-82.7}$ & $ ^{+33.9}_{-20.8}$ & $^{+23.3}_{-3.2}$ &$^{+0.3}_{-0.3}$ &$^{-20.5}_{+24.6}$ \\
\hline 
\multicolumn{7}{|c|}{$d\sigma/dy$ ($P_{e}=+0.33 \pm 0.01$)}\\ 
\hline 
$y$ & $d\sigma/dy$ (pb) & $\delta_{{\rm stat}}$ (\%)&  $\delta_{{\rm syst}}$ (\%)&   $\delta_{{\rm unc}}$ (\%)& $\delta_{\rm trk}$ (\%)& $\delta_{\rm es}$ (\%)\\  
\hline 
$0.05$ & $103.9$ & $ ^{+5.2} _{-4.9}$ & $ ^{+1.5}_{-1.8}$ & $^{+0.6}_{-1.0}$ & $^{+0.7}_{-0.6}$ &$^{+1.2}_{-1.4}$ \\
$0.15$ & $87.0$ & $ ^{+4.5} _{-4.3}$ & $ ^{+1.1}_{-1.2}$ & $^{+0.5}_{-0.8}$ & $^{+0.8}_{-0.8}$ &$^{+0.4}_{-0.4}$ \\
$0.27$ & $66.5$ & $ ^{+4.4} _{-4.2}$ & $ ^{+1.3}_{-1.5}$ & $^{+0.7}_{-1.0}$ & $^{+1.1}_{-1.1}$ &$^{-0.0}_{+0.1}$ \\
$0.41$ & $49.3$ & $ ^{+5.5} _{-5.2}$ & $ ^{+1.8}_{-1.8}$ & $^{+0.3}_{-0.8}$ & $^{+1.4}_{-1.4}$ &$^{-0.9}_{+1.1}$ \\
$0.55$ & $35.6$ & $ ^{+7.0} _{-6.5}$ & $ ^{+2.5}_{-3.1}$ & $^{+0.5}_{-2.0}$ & $^{+1.6}_{-1.5}$ &$^{-1.8}_{+1.8}$ \\
$0.69$ & $25.9$ & $ ^{+9.2} _{-8.5}$ & $ ^{+4.1}_{-4.4}$ & $^{+0.8}_{-2.8}$ & $^{+1.8}_{-1.8}$ &$^{-2.9}_{+3.6}$ \\
$0.83$ & $19.5$ & $ ^{+13.3} _{-11.8}$ & $ ^{+7.7}_{-7.8}$ & $^{+0.3}_{-3.3}$ & $^{+1.9}_{-1.8}$ &$^{-6.8}_{+7.4}$ \\
\hline 
\end{tabular}

\caption{Values of the differential cross-sections $d \sigma /dQ^{2}$,
  $d \sigma /dx$ and $d \sigma /dy$ for $P_{e}=+0.33 \pm 0.01$. The following quantities are given: the value at which the cross section is quoted; the measured cross section; the statistical uncertainty; the total systematic uncertainty ($\delta_{syst}$); the uncorrelated systematic uncertainty ($\delta_{unc}$); the uncertainty on FLT tracking efficiency ($\delta_{trk}$) and the calorimeter energy-scale uncertainty ($\delta_{es}$). Both $\delta_{trk}$ and $\delta_{es}$ have significant correlations between cross-section bins.}  
\label{tab-uncorr_single_pos}
\end{center}
\end{table}

\begin{table}[p]
\begin{center}

\begin{tabular}{|c|c|c|c|c|c|c|} 
\hline 
\multicolumn{7}{|c|}{$d\sigma/dQ^{2}$ ($P_{e}=-0.36 \pm 0.01$)}\\ 
\hline 
$Q^2$ ($\gev^2$)     & $d\sigma/dQ^{2}$ (pb/$\gev^2$)& $\delta_{{\rm stat}}$ (\%)&  $\delta_{{\rm syst}}$ (\%)&   $\delta_{{\rm unc}}$ (\%)& $\delta_{\rm trk}$ (\%)& $\delta_{\rm es}$(\%)\\ 
\hline 
 $ 280$ & $2.25\cdot 10^{-2}$ & $ ^{+ 10.3} _{-9.4}$ & $ ^{+4.1}_{-4.5}$ & $^{+0.6}_{-2.0}$ &$^{+1.7}_{-1.7}$ &$^{+3.7}_{-3.7}$ \\
 $ 530$ & $1.25\cdot 10^{-2}$ & $ ^{+ 9.8} _{-9.0}$ & $ ^{+3.2}_{-3.0}$ & $^{+0.7}_{-0.8}$ &$^{+1.6}_{-1.5}$ &$^{+2.7}_{-2.5}$ \\
 $ 950$ & $8.45\cdot 10^{-3}$ & $ ^{+ 8.3} _{-7.7}$ & $ ^{+2.0}_{-2.4}$ & $^{+0.4}_{-1.5}$ &$^{+1.4}_{-1.4}$ &$^{+1.3}_{-1.3}$ \\
 $ 1700$ & $4.18\cdot 10^{-3}$ & $ ^{+ 8.6} _{-7.9}$ & $ ^{+1.6}_{-1.5}$ & $^{+0.8}_{-0.8}$ &$^{+1.3}_{-1.3}$ &$^{+0.3}_{-0.0}$ \\
 $ 3000$ & $1.97\cdot 10^{-3}$ & $ ^{+ 9.3} _{-8.5}$ & $ ^{+2.8}_{-2.8}$ & $^{+1.1}_{-1.3}$ &$^{+1.2}_{-1.2}$ &$^{-2.2}_{+2.3}$ \\
 $ 5300$ & $6.81\cdot 10^{-4}$ & $ ^{+ 12.0} _{-10.8}$ & $ ^{+5.7}_{-5.6}$ & $^{+0.7}_{-2.1}$ &$^{+1.1}_{-1.1}$ &$^{-5.1}_{+5.5}$ \\
 $ 9500$ & $9.66\cdot 10^{-5}$ & $ ^{+ 24.9} _{-20.2}$ & $ ^{+11.4}_{-8.7}$ & $^{+1.8}_{-0.9}$ &$^{+1.1}_{-1.1}$ &$^{-8.6}_{+11.2}$ \\
 $ 17000$ & $1.80\cdot 10^{-5}$ & $ ^{+ 42.7} _{-31.0}$ & $ ^{+19.2}_{-15.8}$ & $^{+0.0}_{-4.7}$ &$^{+1.1}_{-1.1}$ &$^{-15.1}_{+19.2}$ \\
 $ 30000$ & $1.33\cdot 10^{-6}$ & $ ^{+ 131.9} _{-64.6}$ & $ ^{+32.8}_{-27.4}$ & $^{+0.0}_{-5.5}$ &$^{+1.2}_{-1.2}$ &$^{-26.8}_{+32.8}$ \\
\hline 
\multicolumn{7}{|c|}{$d\sigma/dx$ ($P_{e}=-0.36 \pm 0.01$)}\\ 
\hline 
$x$ & $d\sigma/dx$ (pb)& $\delta_{{\rm stat}}$ (\%)&  $\delta_{{\rm syst}}$ (\%)&   $\delta_{{\rm unc}}$ (\%)& $\delta_{\rm trk}$ (\%)& $\delta_{\rm es}$ (\%)\\ 
\hline 
 $ 0.0078$ & $3.64\cdot 10^{2}$ & $ ^{+ 26.9} _{-21.6}$ & $ ^{+6.9}_{-10.0}$ & $^{+0.7}_{-7.9}$ &$^{+3.1}_{-3.0}$ &$^{+5.9}_{-5.4}$ \\
 $ 0.015$ & $3.32\cdot 10^{2}$ & $ ^{+ 10.8} _{-9.8}$ & $ ^{+4.1}_{-4.5}$ & $^{+1.1}_{-2.8}$ &$^{+2.4}_{-2.3}$ &$^{+3.2}_{-2.7}$ \\
 $ 0.032$ & $1.98\cdot 10^{2}$ & $ ^{+ 7.7} _{-7.2}$ & $ ^{+2.1}_{-2.2}$ & $^{+0.5}_{-0.8}$ &$^{+1.7}_{-1.7}$ &$^{+1.1}_{-1.2}$ \\
 $ 0.068$ & $1.07\cdot 10^{2}$ & $ ^{+ 6.7} _{-6.3}$ & $ ^{+1.4}_{-1.5}$ & $^{+0.5}_{-0.8}$ &$^{+1.2}_{-1.2}$ &$^{-0.3}_{+0.4}$ \\
 $ 0.130$ & $4.87\cdot 10^{1}$ & $ ^{+ 8.6} _{-8.0}$ & $ ^{+2.4}_{-2.3}$ & $^{+0.8}_{-1.2}$ &$^{+0.9}_{-0.9}$ &$^{-1.8}_{+2.0}$ \\
 $ 0.240$ & $1.49\cdot 10^{1}$ & $ ^{+ 12.9} _{-11.5}$ & $ ^{+4.8}_{-4.4}$ & $^{+2.2}_{-1.6}$ &$^{+0.7}_{-0.7}$ &$^{-4.1}_{+4.1}$ \\
 $ 0.420$ & $2.83\cdot 10^{0}$ & $ ^{+ 28.6} _{-22.7}$ & $ ^{+9.7}_{-8.3}$ & $^{+3.6}_{-1.7}$ &$^{+0.5}_{-0.5}$ &$^{-8.1}_{+9.0}$ \\
 $ 0.650$ & $2.35\cdot 10^{-1}$ & $ ^{+ 229.9} _{-82.7}$ & $ ^{+24.5}_{-21.8}$ & $^{+0.0}_{-7.6}$ &$^{+0.4}_{-0.4}$ &$^{-20.4}_{+24.5}$ \\
\hline 
\multicolumn{7}{|c|}{$d\sigma/dy$ ($P_{e}=-0.36 \pm 0.01$)}\\ 
\hline 
$y$ & $d\sigma/dy$ (pb) & $\delta_{{\rm stat}}$ (\%)&  $\delta_{{\rm syst}}$ (\%)&   $\delta_{{\rm unc}}$ (\%)& $\delta_{\rm trk}$ (\%)& $\delta_{\rm es}$ (\%)\\  
\hline 
$0.05$ & $56.2$ & $ ^{+8.4} _{-7.8}$ & $ ^{+1.7}_{-2.0}$ & $^{+0.9}_{-1.2}$ & $^{+0.9}_{-0.9}$ &$^{+1.2}_{-1.4}$ \\
$0.15$ & $39.6$ & $ ^{+7.9} _{-7.4}$ & $ ^{+1.5}_{-1.4}$ & $^{+0.9}_{-0.8}$ & $^{+1.1}_{-1.1}$ &$^{+0.4}_{-0.5}$ \\
$0.27$ & $31.9$ & $ ^{+7.6} _{-7.1}$ & $ ^{+1.5}_{-1.6}$ & $^{+0.7}_{-0.8}$ & $^{+1.4}_{-1.3}$ &$^{-0.0}_{+0.1}$ \\
$0.41$ & $20.5$ & $ ^{+10.3} _{-9.4}$ & $ ^{+2.3}_{-2.1}$ & $^{+1.3}_{-1.0}$ & $^{+1.6}_{-1.6}$ &$^{-0.9}_{+1.1}$ \\
$0.55$ & $18.5$ & $ ^{+11.7} _{-10.5}$ & $ ^{+2.6}_{-3.0}$ & $^{+0.3}_{-1.7}$ & $^{+1.8}_{-1.8}$ &$^{-1.8}_{+1.9}$ \\
$0.69$ & $11.1$ & $ ^{+17.5} _{-15.0}$ & $ ^{+4.3}_{-4.9}$ & $^{+0.9}_{-3.4}$ & $^{+2.2}_{-2.1}$ &$^{-2.9}_{+3.6}$ \\
$0.83$ & $10.5$ & $ ^{+22.7} _{-18.8}$ & $ ^{+8.2}_{-8.3}$ & $^{+1.9}_{-3.9}$ & $^{+2.5}_{-2.4}$ &$^{-6.9}_{+7.6}$ \\
\hline 
\end{tabular}

\caption{Values of the differential cross-sections $d \sigma /dQ^{2}$,
  $d \sigma /dx$ and $d \sigma /dy$ for $P_{e}=-0.36 \pm 0.01$. The following quantities are given: the value at which the cross section is quoted; the measured cross section; the statistical uncertainty; the total systematic uncertainty ($\delta_{syst}$); the uncorrelated systematic uncertainty ($\delta_{unc}$); the uncertainty on FLT tracking efficiency ($\delta_{trk}$) and the calorimeter energy-scale uncertainty ($\delta_{es}$). Both $\delta_{trk}$ and $\delta_{es}$ have significant correlations between cross-section bins.}
\label{tab-uncorr_single_neg}
\end{center}
\end{table}

\begin{table}[p]
\footnotesize
\begin{center}

\begin{tabular}{|c|c|l|l|l|} 
\hline 
$Q^2$ ($\gev^2$) & $x$ & \multicolumn{3}{c|}{$\tilde{\sigma}$} \\ 
\hline 
& & $P_{e}=-0.36$ & $P_{e}=+0.33$ &  $P_{e}=0$\\ 
\hline 
$280$ & $0.0078$ & $(8.23$ $^{+2.84}_{-2.18}$ $^{+0.59}_{-0.76}) \cdot 10^{-1}$ & $(1.44$ $^{+0.31}_{-0.26}$ $^{+0.10}_{-0.12}) \cdot 10^{0}$ & $(1.14$ $^{+0.20}_{-0.17}$ $^{+0.08}_{-0.10}) \cdot 10^{0}$ \\ 
 $280$ & $0.015$ & $(9.07$ $^{+1.72}_{-1.47}$ $^{+0.49}_{-0.51}) \cdot 10^{-1}$ & $(1.85$ $^{+0.20}_{-0.18}$ $^{+0.10}_{-0.11}) \cdot 10^{0}$ & $(1.40$ $^{+0.13}_{-0.12}$ $^{+0.07}_{-0.08}) \cdot 10^{0}$ \\ 
 $280$ & $0.032$ & $(6.39$ $^{+1.27}_{-1.08}$ $^{+0.19}_{-0.22}) \cdot 10^{-1}$ & $(1.12$ $^{+0.14}_{-0.12}$ $^{+0.03}_{-0.04}) \cdot 10^{0}$ & $(8.84$ $^{+0.89}_{-0.81}$ $^{+0.26}_{-0.30}) \cdot 10^{-1}$ \\ 
 $280$ & $0.068$ & $(3.91$ $^{+1.08}_{-0.87}$ $^{+0.11}_{-0.14}) \cdot 10^{-1}$ & $(7.03$ $^{+1.17}_{-1.01}$ $^{+0.20}_{-0.26}) \cdot 10^{-1}$ & $(5.52$ $^{+0.75}_{-0.67}$ $^{+0.16}_{-0.20}) \cdot 10^{-1}$ \\ 
 $280$ & $0.130$ & $(3.27$ $^{+2.58}_{-1.56}$ $^{+0.12}_{-0.12}) \cdot 10^{-1}$ & $(7.88$ $^{+2.85}_{-2.16}$ $^{+0.26}_{-0.28}) \cdot 10^{-1}$ & $(5.74$ $^{+1.76}_{-1.38}$ $^{+0.19}_{-0.21}) \cdot 10^{-1}$ \\ 
 $530$ & $0.0078$ & $(4.86$ $^{+2.90}_{-1.93}$ $^{+0.38}_{-0.56}) \cdot 10^{-1}$ & $(9.61$ $^{+3.18}_{-2.45}$ $^{+0.62}_{-1.35}) \cdot 10^{-1}$ & $(7.37$ $^{+1.98}_{-1.59}$ $^{+0.50}_{-0.95}) \cdot 10^{-1}$ \\ 
 $530$ & $0.015$ & $(6.19$ $^{+1.25}_{-1.06}$ $^{+0.24}_{-0.22}) \cdot 10^{-1}$ & $(1.32$ $^{+0.15}_{-0.13}$ $^{+0.05}_{-0.05}) \cdot 10^{0}$ & $(9.90$ $^{+0.93}_{-0.86}$ $^{+0.37}_{-0.35}) \cdot 10^{-1}$ \\ 
 $530$ & $0.032$ & $(4.63$ $^{+0.89}_{-0.76}$ $^{+0.12}_{-0.12}) \cdot 10^{-1}$ & $(1.55$ $^{+0.13}_{-0.12}$ $^{+0.04}_{-0.04}) \cdot 10^{0}$ & $(1.05$ $^{+0.08}_{-0.07}$ $^{+0.03}_{-0.02}) \cdot 10^{0}$ \\ 
 $530$ & $0.068$ & $(4.61$ $^{+0.86}_{-0.74}$ $^{+0.14}_{-0.11}) \cdot 10^{-1}$ & $(9.04$ $^{+0.98}_{-0.89}$ $^{+0.27}_{-0.21}) \cdot 10^{-1}$ & $(6.93$ $^{+0.63}_{-0.58}$ $^{+0.21}_{-0.16}) \cdot 10^{-1}$ \\ 
 $530$ & $0.130$ & $(1.64$ $^{+0.81}_{-0.57}$ $^{+0.04}_{-0.04}) \cdot 10^{-1}$ & $(5.52$ $^{+1.08}_{-0.92}$ $^{+0.14}_{-0.12}) \cdot 10^{-1}$ & $(3.75$ $^{+0.65}_{-0.56}$ $^{+0.10}_{-0.08}) \cdot 10^{-1}$ \\ 
 $950$ & $0.015$ & $(3.98$ $^{+0.99}_{-0.81}$ $^{+0.15}_{-0.24}) \cdot 10^{-1}$ & $(9.15$ $^{+1.18}_{-1.05}$ $^{+0.26}_{-0.50}) \cdot 10^{-1}$ & $(6.73$ $^{+0.75}_{-0.68}$ $^{+0.20}_{-0.37}) \cdot 10^{-1}$ \\ 
 $950$ & $0.032$ & $(4.30$ $^{+0.69}_{-0.60}$ $^{+0.10}_{-0.12}) \cdot 10^{-1}$ & $(1.04$ $^{+0.09}_{-0.08}$ $^{+0.02}_{-0.03}) \cdot 10^{0}$ & $(7.57$ $^{+0.55}_{-0.51}$ $^{+0.17}_{-0.20}) \cdot 10^{-1}$ \\ 
 $950$ & $0.068$ & $(4.37$ $^{+0.66}_{-0.58}$ $^{+0.07}_{-0.08}) \cdot 10^{-1}$ & $(6.75$ $^{+0.67}_{-0.62}$ $^{+0.10}_{-0.10}) \cdot 10^{-1}$ & $(5.55$ $^{+0.45}_{-0.42}$ $^{+0.09}_{-0.08}) \cdot 10^{-1}$ \\ 
 $950$ & $0.130$ & $(3.04$ $^{+0.69}_{-0.57}$ $^{+0.05}_{-0.06}) \cdot 10^{-1}$ & $(5.98$ $^{+0.77}_{-0.69}$ $^{+0.10}_{-0.11}) \cdot 10^{-1}$ & $(4.58$ $^{+0.50}_{-0.45}$ $^{+0.07}_{-0.08}) \cdot 10^{-1}$ \\ 
 $950$ & $0.240$ & $(1.12$ $^{+0.67}_{-0.44}$ $^{+0.01}_{-0.01}) \cdot 10^{-1}$ & $(2.31$ $^{+0.71}_{-0.56}$ $^{+0.01}_{-0.02}) \cdot 10^{-1}$ & $(1.75$ $^{+0.45}_{-0.36}$ $^{+0.01}_{-0.01}) \cdot 10^{-1}$ \\ 
 $1700$ & $0.032$ & $(3.12$ $^{+0.52}_{-0.45}$ $^{+0.09}_{-0.07}) \cdot 10^{-1}$ & $(7.20$ $^{+0.64}_{-0.59}$ $^{+0.13}_{-0.18}) \cdot 10^{-1}$ & $(5.29$ $^{+0.41}_{-0.38}$ $^{+0.10}_{-0.11}) \cdot 10^{-1}$ \\ 
 $1700$ & $0.068$ & $(2.48$ $^{+0.42}_{-0.36}$ $^{+0.03}_{-0.04}) \cdot 10^{-1}$ & $(7.10$ $^{+0.57}_{-0.53}$ $^{+0.09}_{-0.10}) \cdot 10^{-1}$ & $(4.98$ $^{+0.35}_{-0.33}$ $^{+0.06}_{-0.06}) \cdot 10^{-1}$ \\ 
 $1700$ & $0.130$ & $(2.68$ $^{+0.52}_{-0.44}$ $^{+0.03}_{-0.04}) \cdot 10^{-1}$ & $(3.66$ $^{+0.50}_{-0.45}$ $^{+0.04}_{-0.03}) \cdot 10^{-1}$ & $(3.14$ $^{+0.34}_{-0.31}$ $^{+0.03}_{-0.03}) \cdot 10^{-1}$ \\ 
 $1700$ & $0.240$ & $(1.65$ $^{+0.46}_{-0.37}$ $^{+0.05}_{-0.03}) \cdot 10^{-1}$ & $(2.66$ $^{+0.46}_{-0.40}$ $^{+0.02}_{-0.04}) \cdot 10^{-1}$ & $(2.16$ $^{+0.30}_{-0.27}$ $^{+0.03}_{-0.03}) \cdot 10^{-1}$ \\ 
 $1700$ & $0.420$ & $(1.80$ $^{+4.14}_{-1.49}$ $^{+0.08}_{-0.08}) \cdot 10^{-2}$ & $(9.47$ $^{+5.10}_{-3.49}$ $^{+0.32}_{-0.44}) \cdot 10^{-2}$ & $(6.03$ $^{+2.97}_{-2.09}$ $^{+0.22}_{-0.27}) \cdot 10^{-2}$ \\ 
 $3000$ & $0.032$ & $(3.10$ $^{+0.79}_{-0.64}$ $^{+0.14}_{-0.19}) \cdot 10^{-1}$ & $(4.73$ $^{+0.79}_{-0.68}$ $^{+0.23}_{-0.17}) \cdot 10^{-1}$ & $(3.91$ $^{+0.52}_{-0.46}$ $^{+0.18}_{-0.15}) \cdot 10^{-1}$ \\ 
 $3000$ & $0.068$ & $(2.47$ $^{+0.38}_{-0.33}$ $^{+0.06}_{-0.08}) \cdot 10^{-1}$ & $(5.24$ $^{+0.44}_{-0.41}$ $^{+0.12}_{-0.15}) \cdot 10^{-1}$ & $(3.93$ $^{+0.28}_{-0.27}$ $^{+0.09}_{-0.12}) \cdot 10^{-1}$ \\ 
 $3000$ & $0.130$ & $(2.08$ $^{+0.39}_{-0.34}$ $^{+0.05}_{-0.03}) \cdot 10^{-1}$ & $(3.41$ $^{+0.41}_{-0.37}$ $^{+0.07}_{-0.05}) \cdot 10^{-1}$ & $(2.75$ $^{+0.27}_{-0.25}$ $^{+0.06}_{-0.04}) \cdot 10^{-1}$ \\ 
 $3000$ & $0.240$ & $(9.08$ $^{+2.87}_{-2.25}$ $^{+0.51}_{-0.31}) \cdot 10^{-2}$ & $(2.63$ $^{+0.37}_{-0.33}$ $^{+0.07}_{-0.07}) \cdot 10^{-1}$ & $(1.84$ $^{+0.23}_{-0.21}$ $^{+0.06}_{-0.05}) \cdot 10^{-1}$ \\ 
 $3000$ & $0.420$ & $(2.95$ $^{+2.33}_{-1.41}$ $^{+0.12}_{-0.20}) \cdot 10^{-2}$ & $(6.47$ $^{+2.46}_{-1.84}$ $^{+0.26}_{-0.39}) \cdot 10^{-2}$ & $(4.82$ $^{+1.53}_{-1.19}$ $^{+0.18}_{-0.30}) \cdot 10^{-2}$ \\ 
 $5300$ & $0.068$ & $(1.68$ $^{+0.37}_{-0.31}$ $^{+0.10}_{-0.10}) \cdot 10^{-1}$ & $(3.05$ $^{+0.40}_{-0.35}$ $^{+0.19}_{-0.16}) \cdot 10^{-1}$ & $(2.39$ $^{+0.26}_{-0.23}$ $^{+0.15}_{-0.13}) \cdot 10^{-1}$ \\ 
 $5300$ & $0.130$ & $(1.55$ $^{+0.33}_{-0.27}$ $^{+0.08}_{-0.08}) \cdot 10^{-1}$ & $(2.45$ $^{+0.33}_{-0.29}$ $^{+0.12}_{-0.12}) \cdot 10^{-1}$ & $(2.00$ $^{+0.22}_{-0.20}$ $^{+0.10}_{-0.10}) \cdot 10^{-1}$ \\ 
 $5300$ & $0.240$ & $(9.97$ $^{+2.76}_{-2.21}$ $^{+0.53}_{-0.57}) \cdot 10^{-2}$ & $(1.83$ $^{+0.30}_{-0.26}$ $^{+0.10}_{-0.10}) \cdot 10^{-1}$ & $(1.43$ $^{+0.19}_{-0.17}$ $^{+0.07}_{-0.08}) \cdot 10^{-1}$ \\ 
 $5300$ & $0.420$ & $(2.12$ $^{+1.67}_{-1.01}$ $^{+0.22}_{-0.13}) \cdot 10^{-2}$ & $(1.17$ $^{+0.25}_{-0.21}$ $^{+0.11}_{-0.08}) \cdot 10^{-1}$ & $(7.39$ $^{+1.50}_{-1.26}$ $^{+0.68}_{-0.48}) \cdot 10^{-2}$ \\ 
 $9500$ & $0.130$ & $(4.54$ $^{+2.07}_{-1.48}$ $^{+0.56}_{-0.45}) \cdot 10^{-2}$ & $(1.42$ $^{+0.27}_{-0.23}$ $^{+0.18}_{-0.14}) \cdot 10^{-1}$ & $(9.79$ $^{+1.64}_{-1.42}$ $^{+1.21}_{-0.97}) \cdot 10^{-2}$ \\ 
 $9500$ & $0.240$ & $(3.50$ $^{+1.89}_{-1.29}$ $^{+0.39}_{-0.26}) \cdot 10^{-2}$ & $(1.33$ $^{+0.26}_{-0.22}$ $^{+0.13}_{-0.10}) \cdot 10^{-1}$ & $(8.84$ $^{+1.56}_{-1.34}$ $^{+0.84}_{-0.65}) \cdot 10^{-2}$ \\ 
 $9500$ & $0.420$ & $(3.66$ $^{+1.97}_{-1.35}$ $^{+0.59}_{-0.37}) \cdot 10^{-2}$ & $(4.20$ $^{+1.69}_{-1.25}$ $^{+0.48}_{-0.43}) \cdot 10^{-2}$ & $(3.84$ $^{+1.14}_{-0.90}$ $^{+0.43}_{-0.38}) \cdot 10^{-2}$ \\ 
 $17000$ & $0.240$ & $(3.02$ $^{+1.80}_{-1.20}$ $^{+0.57}_{-0.46}) \cdot 10^{-2}$ & $(3.69$ $^{+1.58}_{-1.14}$ $^{+0.70}_{-0.57}) \cdot 10^{-2}$ & $(3.29$ $^{+1.04}_{-0.82}$ $^{+0.63}_{-0.51}) \cdot 10^{-2}$ \\ 
 $17000$ & $0.420$ & $(1.10$ $^{+1.45}_{-0.71}$ $^{+0.19}_{-0.14}) \cdot 10^{-2}$ & $(3.21$ $^{+1.58}_{-1.11}$ $^{+0.57}_{-0.42}) \cdot 10^{-2}$ & $(2.24$ $^{+0.96}_{-0.70}$ $^{+0.43}_{-0.29}) \cdot 10^{-2}$ \\ 
 $30000$ & $0.420$ & $(5.32$ $^{+12.24}_{-4.41}$ $^{+1.85}_{-1.47}) \cdot 10^{-3}$ & $(1.17$ $^{+1.14}_{-0.64}$ $^{+0.41}_{-0.32}) \cdot 10^{-2}$ & $(8.71$ $^{+6.89}_{-4.17}$ $^{+3.03}_{-2.40}) \cdot 10^{-3}$ \\ 
 \hline 
\end{tabular}

\caption{Values of the reduced cross section. The following quantities
  are given: the values of $Q^2$ and $x$ at which the cross section is
  quoted and the measured cross section, with statistical and systematic uncertainties.}
\label{tab-double}
\end{center}
\end{table}

\begin{table}[p]
\footnotesize
\begin{center}

\begin{tabular}{|c|c|l|c|c|c|c|c|} 
\hline
$Q^2$ ($\gev^2$)     & $x$ &$\tilde{\sigma}$ & $\delta_{{\rm stat}}$ (\%)&  $\delta_{{\rm syst}}$ (\%)&   $\delta_{{\rm unc}}$ (\%)& $\delta_{\rm trk}$ (\%)& $\delta_{\rm es}$ (\%)\\ 
\hline 
$280$ & $0.0078$ & $1.44\cdot 10^{0}$ & $ ^{+ 21.4} _{-17.9}$ & $ ^{+7.0}_{-8.3}$ & $^{+0.5}_{-7.0}$ &$^{+2.3}_{-2.2}$ &$^{+6.5}_{-3.9}$ \\
$280$ & $0.015$ & $1.85\cdot 10^{0}$ & $ ^{+ 10.9} _{-9.9}$ & $ ^{+5.1}_{-5.7}$ & $^{+0.6}_{-3.2}$ &$^{+1.9}_{-1.9}$ &$^{+4.7}_{-4.4}$ \\
$280$ & $0.032$ & $1.12\cdot 10^{0}$ & $ ^{+ 12.2} _{-11.0}$ & $ ^{+2.9}_{-3.3}$ & $^{+0.6}_{-0.4}$ &$^{+1.1}_{-1.1}$ &$^{+2.6}_{-3.1}$ \\
$280$ & $0.068$ & $7.03\cdot 10^{-1}$ & $ ^{+ 16.6} _{-14.4}$ & $ ^{+2.8}_{-3.6}$ & $^{+0.5}_{-1.0}$ &$^{+0.8}_{-0.8}$ &$^{+2.7}_{-3.4}$ \\
$280$ & $0.130$ & $7.88\cdot 10^{-1}$ & $ ^{+ 36.2} _{-27.4}$ & $ ^{+3.3}_{-3.6}$ & $^{+0.5}_{-2.2}$ &$^{+0.5}_{-0.5}$ &$^{+3.2}_{-2.8}$ \\
$530$ & $0.0078$ & $9.61\cdot 10^{-1}$ & $ ^{+ 33.1} _{-25.5}$ & $ ^{+6.4}_{-14.1}$ & $^{+0.9}_{-11.9}$ &$^{+2.6}_{-2.4}$ &$^{+5.7}_{-7.1}$ \\
$530$ & $0.015$ & $1.32\cdot 10^{0}$ & $ ^{+ 11.1} _{-10.0}$ & $ ^{+3.7}_{-3.7}$ & $^{+0.5}_{-1.7}$ &$^{+2.0}_{-1.9}$ &$^{+3.0}_{-2.7}$ \\
$530$ & $0.032$ & $1.55\cdot 10^{0}$ & $ ^{+ 8.3} _{-7.7}$ & $ ^{+2.5}_{-2.3}$ & $^{+0.6}_{-0.4}$ &$^{+1.2}_{-1.2}$ &$^{+2.0}_{-1.9}$ \\
$530$ & $0.068$ & $9.04\cdot 10^{-1}$ & $ ^{+ 10.8} _{-9.9}$ & $ ^{+3.0}_{-2.3}$ & $^{+0.7}_{-0.4}$ &$^{+0.8}_{-0.8}$ &$^{+2.8}_{-2.1}$ \\
$530$ & $0.130$ & $5.52\cdot 10^{-1}$ & $ ^{+ 19.6} _{-16.6}$ & $ ^{+2.6}_{-2.2}$ & $^{+0.8}_{-1.1}$ &$^{+0.6}_{-0.6}$ &$^{+2.4}_{-1.8}$ \\
$950$ & $0.015$ & $9.15\cdot 10^{-1}$ & $ ^{+ 12.9} _{-11.5}$ & $ ^{+2.9}_{-5.5}$ & $^{+0.7}_{-4.9}$ &$^{+2.0}_{-1.9}$ &$^{+1.9}_{-1.3}$ \\
$950$ & $0.032$ & $1.04\cdot 10^{0}$ & $ ^{+ 8.3} _{-7.7}$ & $ ^{+2.3}_{-2.6}$ & $^{+0.6}_{-1.3}$ &$^{+1.4}_{-1.4}$ &$^{+1.6}_{-1.8}$ \\
$950$ & $0.068$ & $6.75\cdot 10^{-1}$ & $ ^{+ 10.0} _{-9.1}$ & $ ^{+1.5}_{-1.4}$ & $^{+0.6}_{-0.9}$ &$^{+0.9}_{-0.8}$ &$^{+1.1}_{-0.7}$ \\
$950$ & $0.130$ & $5.98\cdot 10^{-1}$ & $ ^{+ 12.9} _{-11.5}$ & $ ^{+1.7}_{-1.8}$ & $^{+1.3}_{-1.1}$ &$^{+0.5}_{-0.5}$ &$^{+0.9}_{-1.3}$ \\
$950$ & $0.240$ & $2.31\cdot 10^{-1}$ & $ ^{+ 30.6} _{-24.1}$ & $ ^{+0.4}_{-0.7}$ & $^{+0.0}_{-0.0}$ &$^{+0.4}_{-0.4}$ &$^{-0.0}_{-0.6}$ \\
$1700$ & $0.032$ & $7.20\cdot 10^{-1}$ & $ ^{+ 8.9} _{-8.2}$ & $ ^{+1.8}_{-2.4}$ & $^{+0.5}_{-1.7}$ &$^{+1.7}_{-1.6}$ &$^{+0.2}_{-0.5}$ \\
$1700$ & $0.068$ & $7.10\cdot 10^{-1}$ & $ ^{+ 8.0} _{-7.4}$ & $ ^{+1.2}_{-1.5}$ & $^{+0.7}_{-1.1}$ &$^{+0.9}_{-0.9}$ &$^{+0.3}_{+0.2}$ \\
$1700$ & $0.130$ & $3.66\cdot 10^{-1}$ & $ ^{+ 13.8} _{-12.2}$ & $ ^{+1.1}_{-0.9}$ & $^{+0.6}_{-0.7}$ &$^{+0.6}_{-0.6}$ &$^{+0.6}_{+0.2}$ \\
$1700$ & $0.240$ & $2.66\cdot 10^{-1}$ & $ ^{+ 17.5} _{-15.0}$ & $ ^{+0.8}_{-1.6}$ & $^{+0.4}_{-1.4}$ &$^{+0.5}_{-0.5}$ &$^{-0.7}_{+0.5}$ \\
$1700$ & $0.420$ & $9.47\cdot 10^{-2}$ & $ ^{+ 53.9} _{-36.8}$ & $ ^{+3.4}_{-4.7}$ & $^{+2.0}_{-4.5}$ &$^{+0.2}_{-0.2}$ &$^{-1.2}_{+2.7}$ \\
$3000$ & $0.032$ & $4.73\cdot 10^{-1}$ & $ ^{+ 16.6} _{-14.4}$ & $ ^{+4.8}_{-3.6}$ & $^{+1.8}_{-1.6}$ &$^{+1.8}_{-1.8}$ &$^{-2.8}_{+4.1}$ \\
$3000$ & $0.068$ & $5.24\cdot 10^{-1}$ & $ ^{+ 8.5} _{-7.9}$ & $ ^{+2.2}_{-2.9}$ & $^{+0.4}_{-0.8}$ &$^{+1.2}_{-1.2}$ &$^{-2.6}_{+1.8}$ \\
$3000$ & $0.130$ & $3.41\cdot 10^{-1}$ & $ ^{+ 12.1} _{-10.8}$ & $ ^{+2.0}_{-1.5}$ & $^{+1.0}_{-0.9}$ &$^{+0.7}_{-0.7}$ &$^{-1.0}_{+1.6}$ \\
$3000$ & $0.240$ & $2.63\cdot 10^{-1}$ & $ ^{+ 14.3} _{-12.6}$ & $ ^{+2.6}_{-2.8}$ & $^{+0.4}_{-1.7}$ &$^{+0.4}_{-0.4}$ &$^{-2.2}_{+2.5}$ \\
$3000$ & $0.420$ & $6.47\cdot 10^{-2}$ & $ ^{+ 38.0} _{-28.4}$ & $ ^{+4.0}_{-6.0}$ & $^{+1.7}_{-2.3}$ &$^{+0.3}_{-0.3}$ &$^{-5.5}_{+3.6}$ \\
$5300$ & $0.068$ & $3.05\cdot 10^{-1}$ & $ ^{+ 13.0} _{-11.6}$ & $ ^{+6.2}_{-5.3}$ & $^{+1.0}_{-1.1}$ &$^{+1.6}_{-1.5}$ &$^{-4.9}_{+6.0}$ \\
$5300$ & $0.130$ & $2.45\cdot 10^{-1}$ & $ ^{+ 13.6} _{-12.0}$ & $ ^{+5.0}_{-4.9}$ & $^{+0.8}_{-0.8}$ &$^{+0.8}_{-0.8}$ &$^{-4.8}_{+4.9}$ \\
$5300$ & $0.240$ & $1.83\cdot 10^{-1}$ & $ ^{+ 16.2} _{-14.1}$ & $ ^{+5.2}_{-5.5}$ & $^{+2.0}_{-1.0}$ &$^{+0.6}_{-0.6}$ &$^{-5.4}_{+4.8}$ \\
$5300$ & $0.420$ & $1.17\cdot 10^{-1}$ & $ ^{+ 21.8} _{-18.2}$ & $ ^{+9.2}_{-7.0}$ & $^{+2.3}_{-3.9}$ &$^{+0.4}_{-0.4}$ &$^{-5.8}_{+8.9}$ \\
$9500$ & $0.130$ & $1.42\cdot 10^{-1}$ & $ ^{+ 19.0} _{-16.2}$ & $ ^{+12.6}_{-10.0}$ & $^{+2.5}_{-3.4}$ &$^{+1.3}_{-1.3}$ &$^{-9.3}_{+12.3}$ \\
$9500$ & $0.240$ & $1.33\cdot 10^{-1}$ & $ ^{+ 19.6} _{-16.6}$ & $ ^{+9.5}_{-7.5}$ & $^{+0.6}_{-2.1}$ &$^{+0.7}_{-0.7}$ &$^{-7.2}_{+9.5}$ \\
$9500$ & $0.420$ & $4.20\cdot 10^{-2}$ & $ ^{+ 40.1} _{-29.7}$ & $ ^{+11.3}_{-10.2}$ & $^{+3.2}_{-2.5}$ &$^{+0.4}_{-0.4}$ &$^{-9.8}_{+10.9}$ \\
$17000$ & $0.240$ & $3.69\cdot 10^{-2}$ & $ ^{+ 42.7} _{-31.0}$ & $ ^{+19.0}_{-15.4}$ & $^{+0.0}_{-4.8}$ &$^{+1.0}_{-1.0}$ &$^{-14.6}_{+18.9}$ \\
$17000$ & $0.420$ & $3.21\cdot 10^{-2}$ & $ ^{+ 49.3} _{-34.6}$ & $ ^{+17.7}_{-12.9}$ & $^{+0.0}_{-3.5}$ &$^{+0.6}_{-0.5}$ &$^{-12.4}_{+17.7}$ \\
$30000$ & $0.420$ & $1.17\cdot 10^{-2}$ & $ ^{+ 97.3} _{-54.4}$ & $ ^{+34.8}_{-27.5}$ & $^{+0.0}_{-4.8}$ &$^{+0.9}_{-0.9}$ &$^{-27.1}_{+34.8}$ \\
\hline 
\end{tabular}

\caption{Values of the reduced cross section for $P_{e}=+0.33 \pm 0.01$. The following quantities are given: the values of $Q^2$ and $x$ at which the cross section is quoted; the measured cross section; the statistical uncertainty; the total systematic uncertainty ($\delta_{syst}$); the uncorrelated systematic uncertainty  ($\delta_{unc}$); the uncertainty on FLT tracking efficiency ($\delta_{trk}$) and the calorimeter energy-scale uncertainty ($\delta_{es}$). Both $\delta_{trk}$ and $\delta_{es}$ have significant correlations between cross-section bins.}
\label{tab-uncorr_double_pos}
\end{center}
\end{table}

\begin{table}[p]
\footnotesize
\begin{center}

\begin{tabular}{|c|c|c|c|c|c|c|c|} 
\hline
$Q^2$ ($\gev^2$)     & $x$ &$\tilde{\sigma}$ & $\delta_{{\rm stat}}$ (\%)&  $\delta_{{\rm syst}}$ (\%)&   $\delta_{{\rm unc}}$ (\%)& $\delta_{\rm trk}$ (\%)& $\delta_{\rm es}$ (\%)\\ 
\hline 
$280$ & $0.0078$ & $8.23\cdot 10^{-1}$ & $ ^{+ 34.5} _{-26.4}$ & $ ^{+7.2}_{-9.2}$ & $^{+0.9}_{-7.8}$ &$^{+2.9}_{-2.7}$ &$^{+6.2}_{-4.0}$ \\
$280$ & $0.015$ & $9.07\cdot 10^{-1}$ & $ ^{+ 19.0} _{-16.2}$ & $ ^{+5.3}_{-5.7}$ & $^{+0.5}_{-2.7}$ &$^{+2.3}_{-2.2}$ &$^{+4.7}_{-4.4}$ \\
$280$ & $0.032$ & $6.39\cdot 10^{-1}$ & $ ^{+ 19.9} _{-16.8}$ & $ ^{+3.0}_{-3.5}$ & $^{+0.6}_{-0.6}$ &$^{+1.4}_{-1.4}$ &$^{+2.6}_{-3.1}$ \\
$280$ & $0.068$ & $3.91\cdot 10^{-1}$ & $ ^{+ 27.7} _{-22.1}$ & $ ^{+2.9}_{-3.7}$ & $^{+0.7}_{-0.9}$ &$^{+1.1}_{-1.0}$ &$^{+2.6}_{-3.4}$ \\
$280$ & $0.130$ & $3.27\cdot 10^{-1}$ & $ ^{+ 79.1} _{-47.9}$ & $ ^{+3.7}_{-3.7}$ & $^{+1.5}_{-2.4}$ &$^{+0.7}_{-0.7}$ &$^{+3.2}_{-2.7}$ \\
$530$ & $0.0078$ & $4.86\cdot 10^{-1}$ & $ ^{+ 59.7} _{-39.6}$ & $ ^{+7.7}_{-11.5}$ & $^{+2.4}_{-8.2}$ &$^{+3.7}_{-3.4}$ &$^{+6.1}_{-7.3}$ \\
$530$ & $0.015$ & $6.19\cdot 10^{-1}$ & $ ^{+ 20.3} _{-17.1}$ & $ ^{+3.8}_{-3.5}$ & $^{+0.6}_{-0.5}$ &$^{+2.2}_{-2.1}$ &$^{+3.0}_{-2.7}$ \\
$530$ & $0.032$ & $4.63\cdot 10^{-1}$ & $ ^{+ 19.3} _{-16.4}$ & $ ^{+2.7}_{-2.5}$ & $^{+0.9}_{-0.6}$ &$^{+1.5}_{-1.4}$ &$^{+2.0}_{-2.0}$ \\
$530$ & $0.068$ & $4.61\cdot 10^{-1}$ & $ ^{+ 18.7} _{-15.9}$ & $ ^{+3.1}_{-2.5}$ & $^{+0.5}_{-0.7}$ &$^{+1.0}_{-1.0}$ &$^{+2.8}_{-2.1}$ \\
$530$ & $0.130$ & $1.64\cdot 10^{-1}$ & $ ^{+ 49.3} _{-34.6}$ & $ ^{+2.7}_{-2.4}$ & $^{+0.9}_{-1.3}$ &$^{+0.8}_{-0.8}$ &$^{+2.4}_{-1.8}$ \\
$950$ & $0.015$ & $3.98\cdot 10^{-1}$ & $ ^{+ 24.9} _{-20.3}$ & $ ^{+3.9}_{-6.0}$ & $^{+2.3}_{-5.4}$ &$^{+2.5}_{-2.4}$ &$^{+1.8}_{-1.3}$ \\
$950$ & $0.032$ & $4.30\cdot 10^{-1}$ & $ ^{+ 16.1} _{-14.0}$ & $ ^{+2.3}_{-2.7}$ & $^{+0.2}_{-1.3}$ &$^{+1.7}_{-1.6}$ &$^{+1.6}_{-1.8}$ \\
$950$ & $0.068$ & $4.37\cdot 10^{-1}$ & $ ^{+ 15.1} _{-13.2}$ & $ ^{+1.6}_{-1.7}$ & $^{+0.4}_{-1.2}$ &$^{+1.1}_{-1.1}$ &$^{+1.2}_{-0.7}$ \\
$950$ & $0.130$ & $3.04\cdot 10^{-1}$ & $ ^{+ 22.7} _{-18.8}$ & $ ^{+1.6}_{-2.0}$ & $^{+1.1}_{-1.4}$ &$^{+0.7}_{-0.7}$ &$^{+0.9}_{-1.3}$ \\
$950$ & $0.240$ & $1.12\cdot 10^{-1}$ & $ ^{+ 59.7} _{-39.6}$ & $ ^{+0.6}_{-0.9}$ & $^{+0.0}_{-0.0}$ &$^{+0.6}_{-0.6}$ &$^{-0.0}_{-0.7}$ \\
$1700$ & $0.032$ & $3.12\cdot 10^{-1}$ & $ ^{+ 16.8} _{-14.5}$ & $ ^{+2.9}_{-2.2}$ & $^{+2.1}_{-1.1}$ &$^{+2.0}_{-1.9}$ &$^{+0.3}_{-0.5}$ \\
$1700$ & $0.068$ & $2.48\cdot 10^{-1}$ & $ ^{+ 16.8} _{-14.5}$ & $ ^{+1.3}_{-1.4}$ & $^{+0.5}_{-0.9}$ &$^{+1.1}_{-1.1}$ &$^{+0.3}_{+0.2}$ \\
$1700$ & $0.130$ & $2.68\cdot 10^{-1}$ & $ ^{+ 19.6} _{-16.6}$ & $ ^{+1.3}_{-1.6}$ & $^{+0.8}_{-1.4}$ &$^{+0.8}_{-0.8}$ &$^{+0.6}_{+0.2}$ \\
$1700$ & $0.240$ & $1.65\cdot 10^{-1}$ & $ ^{+ 27.7} _{-22.1}$ & $ ^{+3.0}_{-1.8}$ & $^{+2.9}_{-1.6}$ &$^{+0.6}_{-0.6}$ &$^{-0.7}_{+0.5}$ \\
$1700$ & $0.420$ & $1.80\cdot 10^{-2}$ & $ ^{+ 229.9} _{-82.7}$ & $ ^{+4.5}_{-4.2}$ & $^{+3.4}_{-4.0}$ &$^{+0.3}_{-0.3}$ &$^{-1.1}_{+2.9}$ \\
$3000$ & $0.032$ & $3.10\cdot 10^{-1}$ & $ ^{+ 25.5} _{-20.7}$ & $ ^{+4.6}_{-6.0}$ & $^{+0.3}_{-4.9}$ &$^{+2.2}_{-2.1}$ &$^{-2.7}_{+4.1}$ \\
$3000$ & $0.068$ & $2.47\cdot 10^{-1}$ & $ ^{+ 15.2} _{-13.3}$ & $ ^{+2.5}_{-3.1}$ & $^{+1.1}_{-1.1}$ &$^{+1.4}_{-1.4}$ &$^{-2.5}_{+1.8}$ \\
$3000$ & $0.130$ & $2.08\cdot 10^{-1}$ & $ ^{+ 19.0} _{-16.2}$ & $ ^{+2.6}_{-1.6}$ & $^{+1.9}_{-0.9}$ &$^{+0.9}_{-0.9}$ &$^{-1.0}_{+1.6}$ \\
$3000$ & $0.240$ & $9.08\cdot 10^{-2}$ & $ ^{+ 31.6} _{-24.7}$ & $ ^{+5.6}_{-3.4}$ & $^{+5.0}_{-2.5}$ &$^{+0.6}_{-0.6}$ &$^{-2.2}_{+2.5}$ \\
$3000$ & $0.420$ & $2.95\cdot 10^{-2}$ & $ ^{+ 79.1} _{-47.8}$ & $ ^{+3.9}_{-6.8}$ & $^{+1.7}_{-3.7}$ &$^{+0.5}_{-0.5}$ &$^{-5.6}_{+3.5}$ \\
$5300$ & $0.068$ & $1.68\cdot 10^{-1}$ & $ ^{+ 21.8} _{-18.2}$ & $ ^{+6.2}_{-6.1}$ & $^{+0.5}_{-3.2}$ &$^{+1.7}_{-1.6}$ &$^{-4.9}_{+6.0}$ \\
$5300$ & $0.130$ & $1.55\cdot 10^{-1}$ & $ ^{+ 21.0} _{-17.6}$ & $ ^{+5.1}_{-5.3}$ & $^{+0.8}_{-2.1}$ &$^{+1.0}_{-1.0}$ &$^{-4.8}_{+4.9}$ \\
$5300$ & $0.240$ & $9.97\cdot 10^{-2}$ & $ ^{+ 27.7} _{-22.2}$ & $ ^{+5.3}_{-5.7}$ & $^{+2.2}_{-1.6}$ &$^{+0.7}_{-0.7}$ &$^{-5.4}_{+4.8}$ \\
$5300$ & $0.420$ & $2.12\cdot 10^{-2}$ & $ ^{+ 79.1} _{-47.9}$ & $ ^{+10.5}_{-6.3}$ & $^{+5.6}_{-2.6}$ &$^{+0.5}_{-0.5}$ &$^{-5.7}_{+8.9}$ \\
$9500$ & $0.130$ & $4.54\cdot 10^{-2}$ & $ ^{+ 45.7} _{-32.7}$ & $ ^{+12.3}_{-10.0}$ & $^{+0.0}_{-3.0}$ &$^{+1.5}_{-1.5}$ &$^{-9.4}_{+12.3}$ \\
$9500$ & $0.240$ & $3.50\cdot 10^{-2}$ & $ ^{+ 53.9} _{-36.8}$ & $ ^{+11.1}_{-7.6}$ & $^{+5.6}_{-2.2}$ &$^{+0.8}_{-0.8}$ &$^{-7.2}_{+9.5}$ \\
$9500$ & $0.420$ & $3.66\cdot 10^{-2}$ & $ ^{+ 53.9} _{-36.8}$ & $ ^{+16.1}_{-10.1}$ & $^{+11.9}_{-2.4}$ &$^{+0.5}_{-0.5}$ &$^{-9.8}_{+10.8}$ \\
$17000$ & $0.240$ & $3.02\cdot 10^{-2}$ & $ ^{+ 59.7} _{-39.6}$ & $ ^{+19.0}_{-15.3}$ & $^{+0.0}_{-4.7}$ &$^{+1.2}_{-1.1}$ &$^{-14.5}_{+19.0}$ \\
$17000$ & $0.420$ & $1.10\cdot 10^{-2}$ & $ ^{+ 131.8} _{-64.6}$ & $ ^{+17.6}_{-12.9}$ & $^{+0.0}_{-3.5}$ &$^{+0.7}_{-0.7}$ &$^{-12.4}_{+17.6}$ \\
$30000$ & $0.420$ & $5.32\cdot 10^{-3}$ & $ ^{+ 229.9} _{-82.7}$ & $ ^{+34.7}_{-27.6}$ & $^{+0.0}_{-4.7}$ &$^{+1.2}_{-1.1}$ &$^{-27.2}_{+34.7}$ \\
\hline 
\end{tabular}

\caption{Values of the reduced cross section for $P_{e}=-0.36 \pm 0.01$. The following quantities are given: the values of $Q^2$ and $x$ at which the cross section is quoted; the measured cross section; the statistical uncertainty; the total systematic uncertainty ($\delta_{syst}$); the uncorrelated systematic uncertainty  ($\delta_{unc}$); the uncertainty on FLT tracking efficiency ($\delta_{trk}$) and the calorimeter energy-scale uncertainty ($\delta_{es}$). Both $\delta_{trk}$ and $\delta_{es}$ have significant correlations between cross-section bins.}
\label{tab-uncorr_double_neg}
\end{center}
\end{table}

\begin{table}[p]
\begin{center}

\begin{tabular}{|r|c|} 
\hline 
Polarisation & $\sigma^{CC}$ (pb) \\ 
\hline 
$-0.413 \pm 0.016$ & $20.7$ $^{+1.4} _{-1.3}$ $(\textnormal{stat.})$ $\pm 0.5 $ $(\textnormal{lumi.})$ $^{+ 0.3} _{-0.4}$ $(\textnormal{syst.})$ \\ 
$-0.366 \pm 0.015$ & $22.5$ $^{+1.5} _{-1.4}$ $(\textnormal{stat.})$ $\pm 0.6 $ $(\textnormal{lumi.})$ $^{+ 0.4} _{-0.4}$ $(\textnormal{syst.})$ \\ 
$-0.306 \pm 0.012$ & $25.1$ $^{+1.5} _{-1.5}$ $(\textnormal{stat.})$ $\pm 0.7 $ $(\textnormal{lumi.})$ $^{+ 0.4} _{-0.4}$ $(\textnormal{syst.})$ \\ 
$0.259 \pm 0.010$ & $46.4$ $^{+2.0} _{-1.9}$ $(\textnormal{stat.})$ $\pm 1.2 $ $(\textnormal{lumi.})$ $^{+ 0.6} _{-0.7}$ $(\textnormal{syst.})$ \\ 
$0.303 \pm 0.011$ & $46.7$ $^{+2.0} _{-2.0}$ $(\textnormal{stat.})$ $\pm 1.2 $ $(\textnormal{lumi.})$ $^{+ 0.6} _{-0.8}$ $(\textnormal{syst.})$ \\ 
$0.339 \pm 0.013$ & $48.4$ $^{+2.1} _{-2.0}$ $(\textnormal{stat.})$ $\pm 1.3 $ $(\textnormal{lumi.})$ $^{+ 0.6} _{-0.8}$ $(\textnormal{syst.})$ \\ 
$0.416 \pm 0.015$ & $51.4$ $^{+2.1} _{-2.1}$ $(\textnormal{stat.})$ $\pm 1.3 $ $(\textnormal{lumi.})$ $^{+ 0.7} _{-0.8}$ $(\textnormal{syst.})$ \\ 
\hline 
\end{tabular}

\caption{Values of the total cross section, $\sigma^{CC}$, measured at different values of polarisation of the positron beam.  The following quantities are given: the polarisation value at which the cross section is quoted and the measured cross section, with statistical, luminosity and systematic uncertainties.}
\label{tab-polbins}
\end{center}
\end{table}

%% file: DESY-10-129-fig.tex
\begin{figure}[p]
\vfill
\begin{center}
\includegraphics[width=6in]{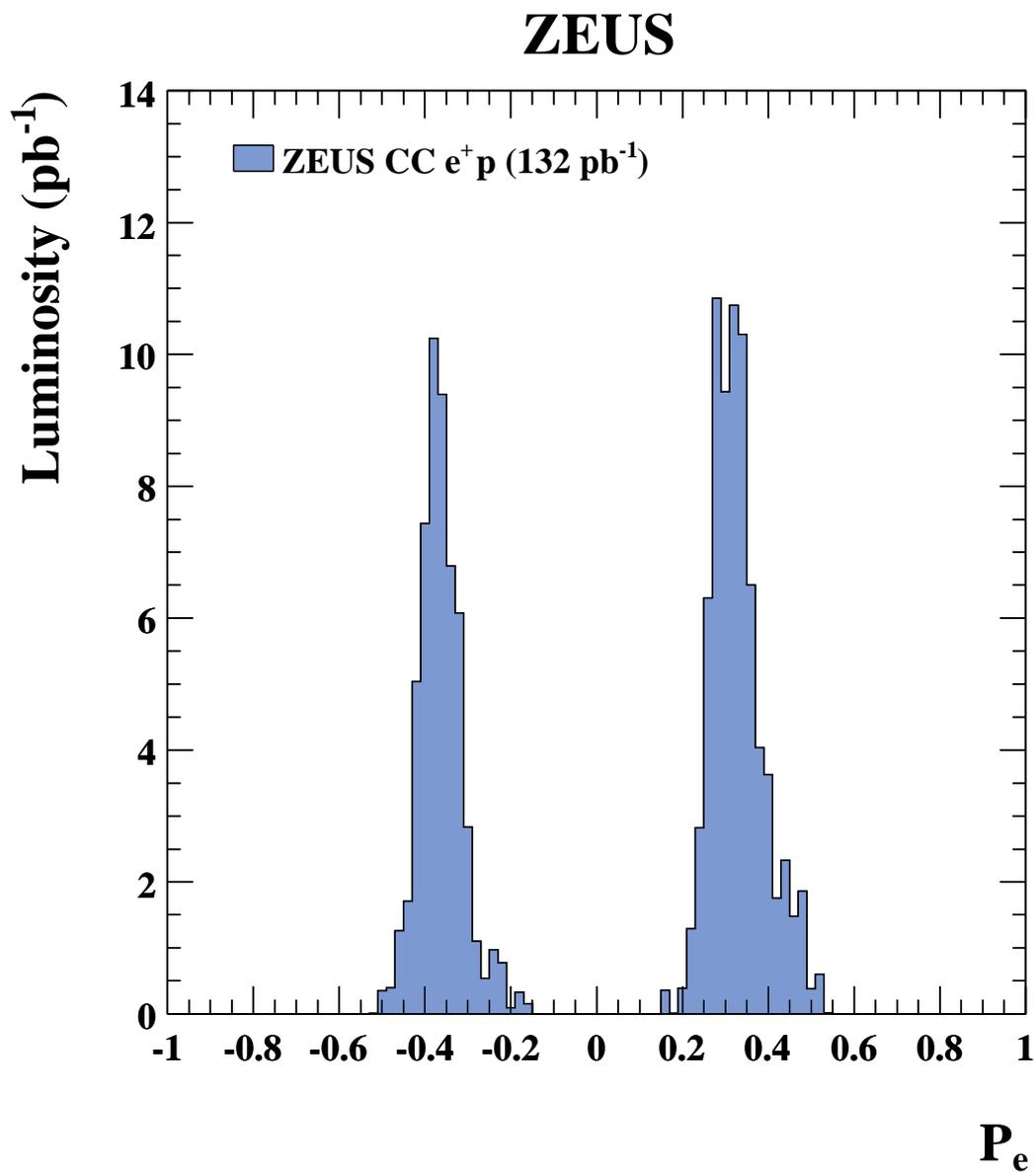}
\end{center}
  \caption{ The integrated luminosity collected as a function of  the
    longitudinal polarisation of the positron beam. Events from runs with 
    mean absolute polarisation less than 15\% were rejected. }
  \label{fig-lumipol}
\vfill
\end{figure}

\begin{figure}[p]
\vfill
\begin{center}
\includegraphics[width=6in]{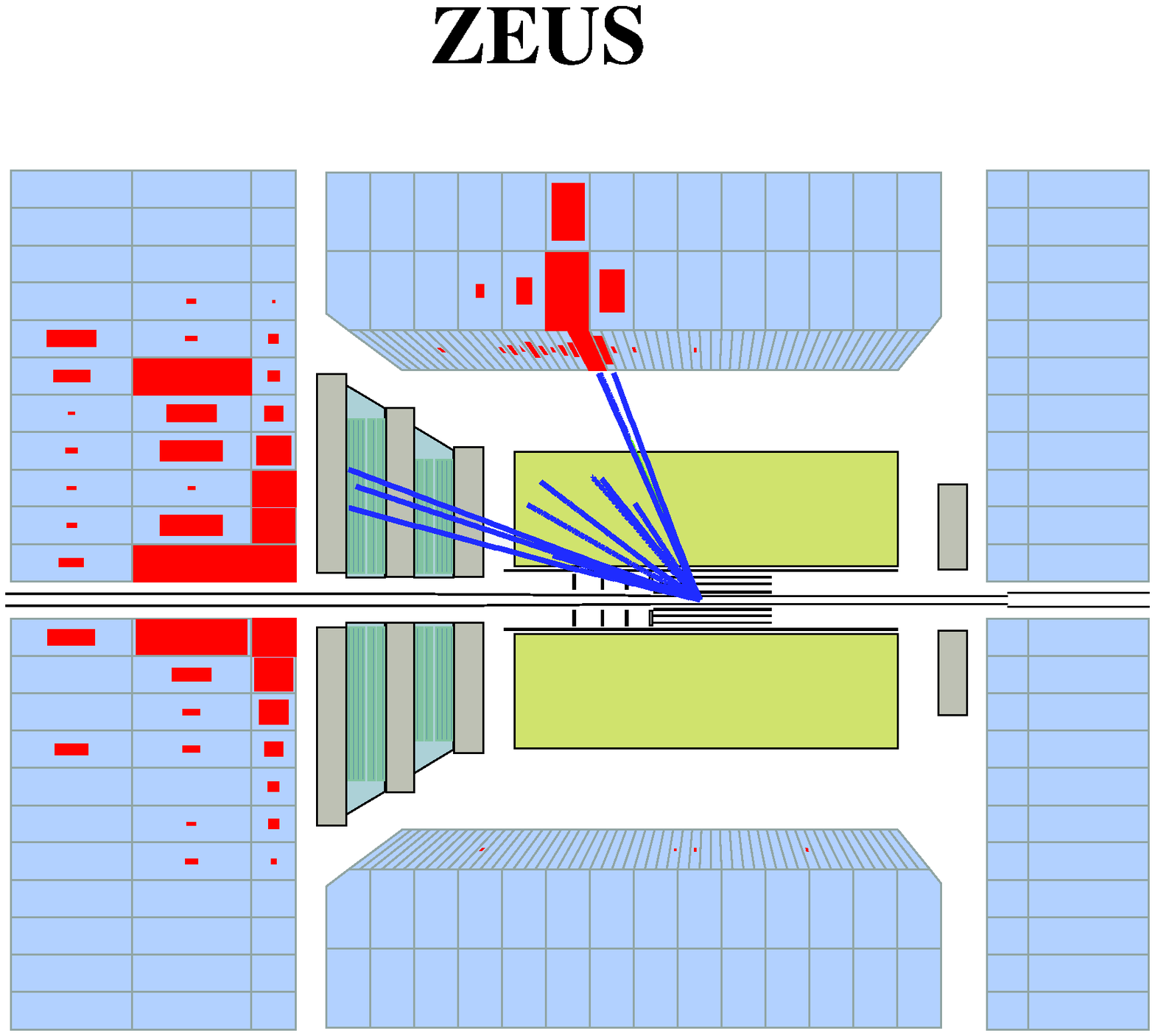}
\end{center}
  \caption{ A charged current event with $Q^2$ = $53\,060$ $GeV^2$ and $x$ = 0.59}
  \label{fig-event}
\vfill
\end{figure}

\begin{figure}[p]
\vfill
  \begin{center}
    \includegraphics[width=6in]{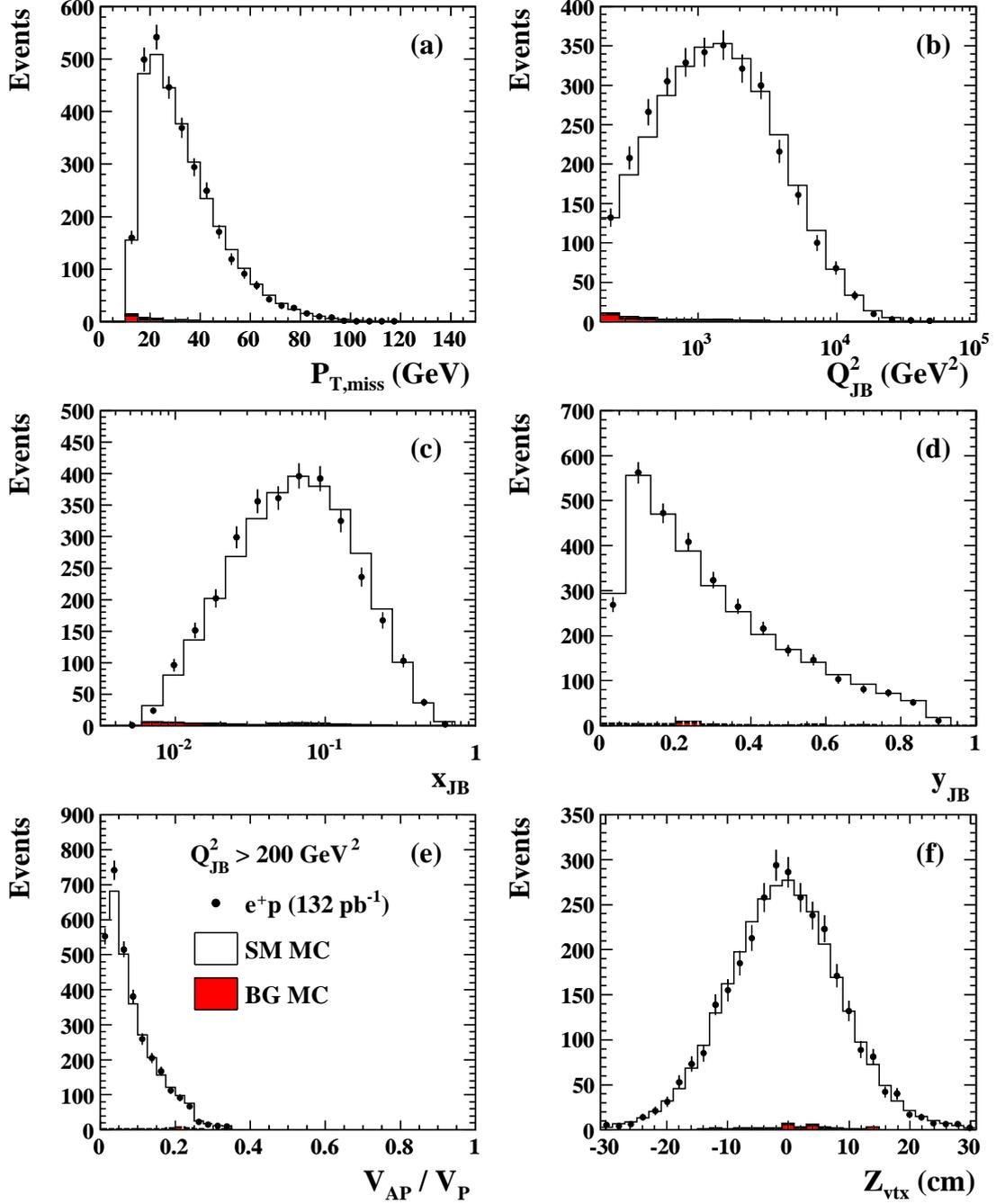}
 \end{center}
  \caption{ Comparison of the total $e^+ p$ CC data sample with the
    expectations of the Monte Carlo simulation described in the text.
    The distributions of (a) $\PTM$, (b)
    $Q^{2}_{\rm JB}$, (c) $x_{\rm JB}$, (d) $y_{\rm JB}$, (e)
    $V_{AP}/V_{P}$ and (f) $Z_{vtx}$, are shown. The points represent data. 
  The open (filled) histograms represent the signal (background) MC.}
  \label{fig-cc_ctrl}
\vfill
\end{figure}

\begin{figure}[p]
\vfill
  \begin{center}
    \includegraphics[width=6in]{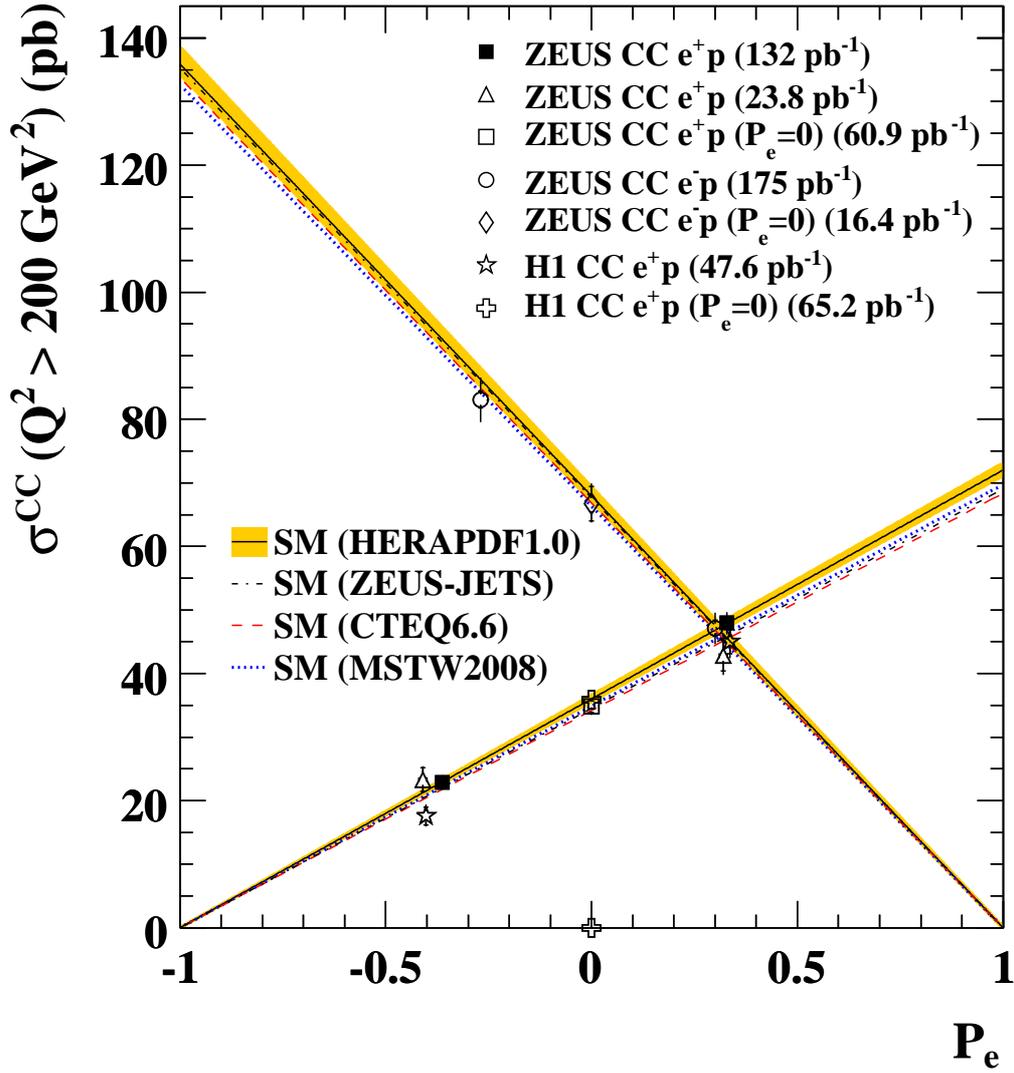}
  \end{center}
  \caption{ The total cross sections for $e^+ p$ (this analysis,
  filled squares) and $e^- p$ CC DIS
    as a function of the longitudinal  polarisation of the lepton
    beam. The lines show the SM predictions obtained with HERAPDF1.0, 
    ZEUS-JETS, CTEQ6.6 and MSTW2008 PDFs. The shaded band shows the 
    total uncertainty from the HERAPDF1.0 PDFs.}
  \label{fig-cctotal}
\vfill
\end{figure}

\begin{figure}
  \begin{center}
    \includegraphics[width=1.0\textwidth]{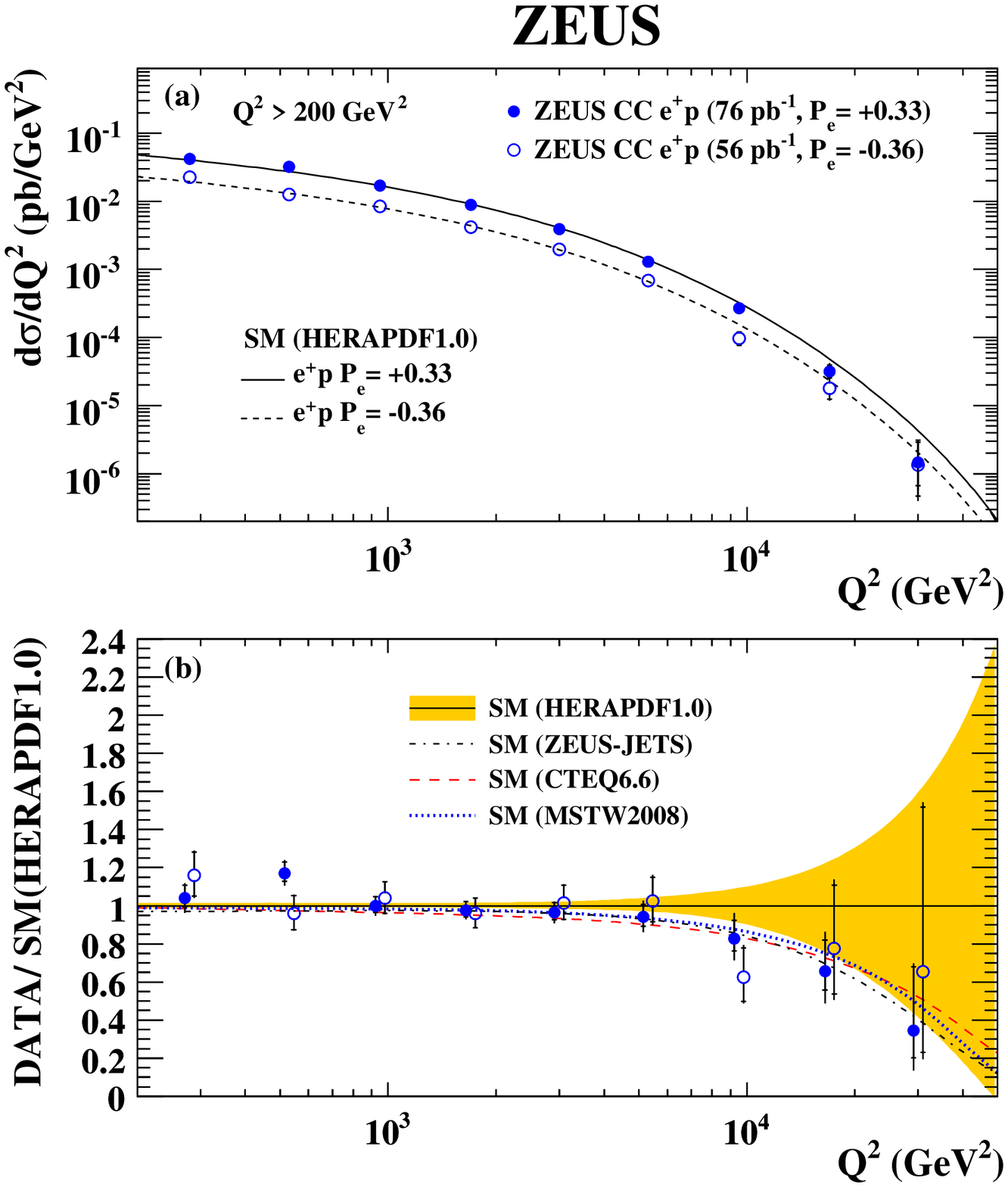}
  \end{center}
  \caption{ (a) The $e^+p$ CC DIS cross-section $d\sigma/dQ^2$ for
    data and the Standard Model expectation evaluated using the
    HERAPDF1.0 PDFs.  The positive (negative) polarisation data are
    shown as the filled (open) points, the statistical uncertainties
    are indicated by the inner error bars (delimited by horizontal
    lines)  and the full error bars show the total uncertainty
    obtained by adding  the statistical and systematic contributions
    in quadrature.  (b) The ratio of the measured cross-section,
    $d\sigma/dQ^2$, to the Standard Model expectation evaluated using
    the HERAPDF1.0 PDFs.  The shaded band shows the total 
    uncertainty from the HERAPDF1.0 PDFs. The curves show the ratio of the
    predictions of the SM evaluated using the ZEUS-JETS, CTEQ6.6  
     and MSTW2008 PDFs to the prediction from the HERAPDF1.0 PDFs.}
  \label{fig-dsdq2}
\end{figure}

\begin{figure}
  \begin{center}
    \includegraphics[width=1.0\textwidth]{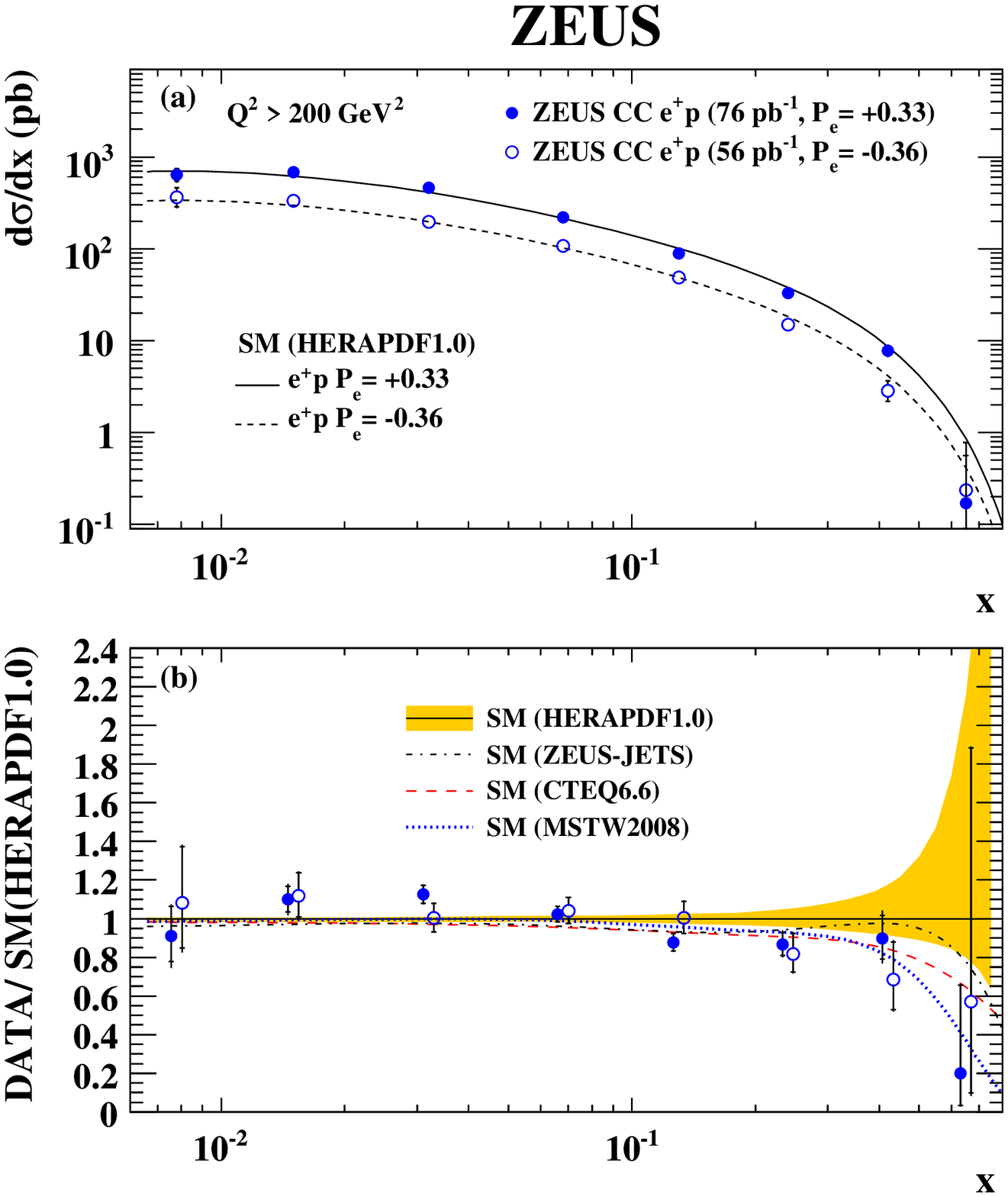}
  \end{center}
  \caption{ (a) The $e^+p$ CC DIS cross-section $d\sigma/dx$ for data
    and the Standard Model expectation evaluated using the HERAPDF1.0
    PDFs.  The positive (negative) polarisation data are shown as the
    filled (open) points, the statistical uncertainties    are
    indicated by the inner error bars (delimited by horizontal lines)
    and the full error bars show the total uncertainty obtained by
    adding  the statistical and systematic contributions in
    quadrature.  (b) The ratio of the measured cross-section,
    $d\sigma/dx$, to the Standard Model expectation evaluated using
    the HERAPDF1.0 PDFs.  The shaded band shows the total
    uncertainty from the HERAPDF1.0 PDFs. The curves show the ratio of the
    predictions of the SM evaluated using the ZEUS-JETS, CTEQ6.6 
     and MSTW2008 PDFs to the prediction from the HERAPDF1.0 PDFs.}
  \label{fig-dsdx}
\end{figure}

\begin{figure}
  \begin{center}
    \includegraphics[width=1.0\textwidth]{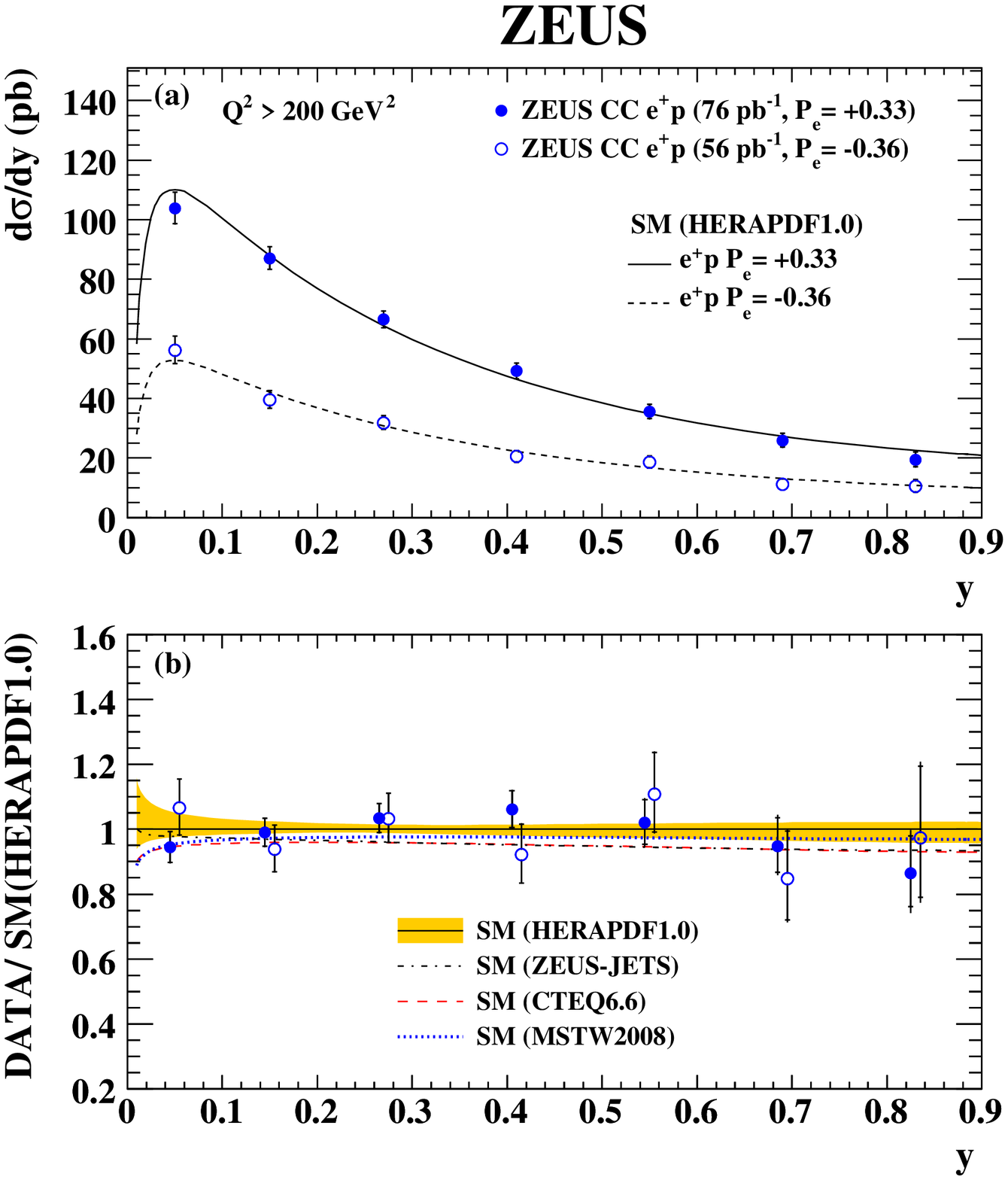}
  \end{center}
  \caption{ (a) The $e^+p$ CC DIS cross-section $d\sigma/dy$ for data
    and the Standard Model expectation evaluated using the HERAPDF1.0 
    PDFs.  The positive (negative) polarisation data are shown as the
    filled (open) points, the statistical uncertainties    are
    indicated by the inner error bars (delimited by horizontal lines)
    and the full error bars show the total uncertainty obtained by
    adding  the statistical and systematic contributions in
    quadrature.  (b) The ratio of the measured cross-section,
    $d\sigma/dy$, to the Standard Model expectation evaluated using
    the HERAPDF1.0 PDFs.  The shaded band shows the total 
    uncertainty from the HERAPDF1.0 PDFs. The curves show the ratio of the
    predictions of the SM evaluated using the ZEUS-JETS, CTEQ6.6  
     and MSTW2008 PDFs to the prediction from the HERAPDF1.0 PDFs.}
  \label{fig-dsdy}
\end{figure}

\begin{figure}
  \begin{center}
    \includegraphics[width=1.0\textwidth]{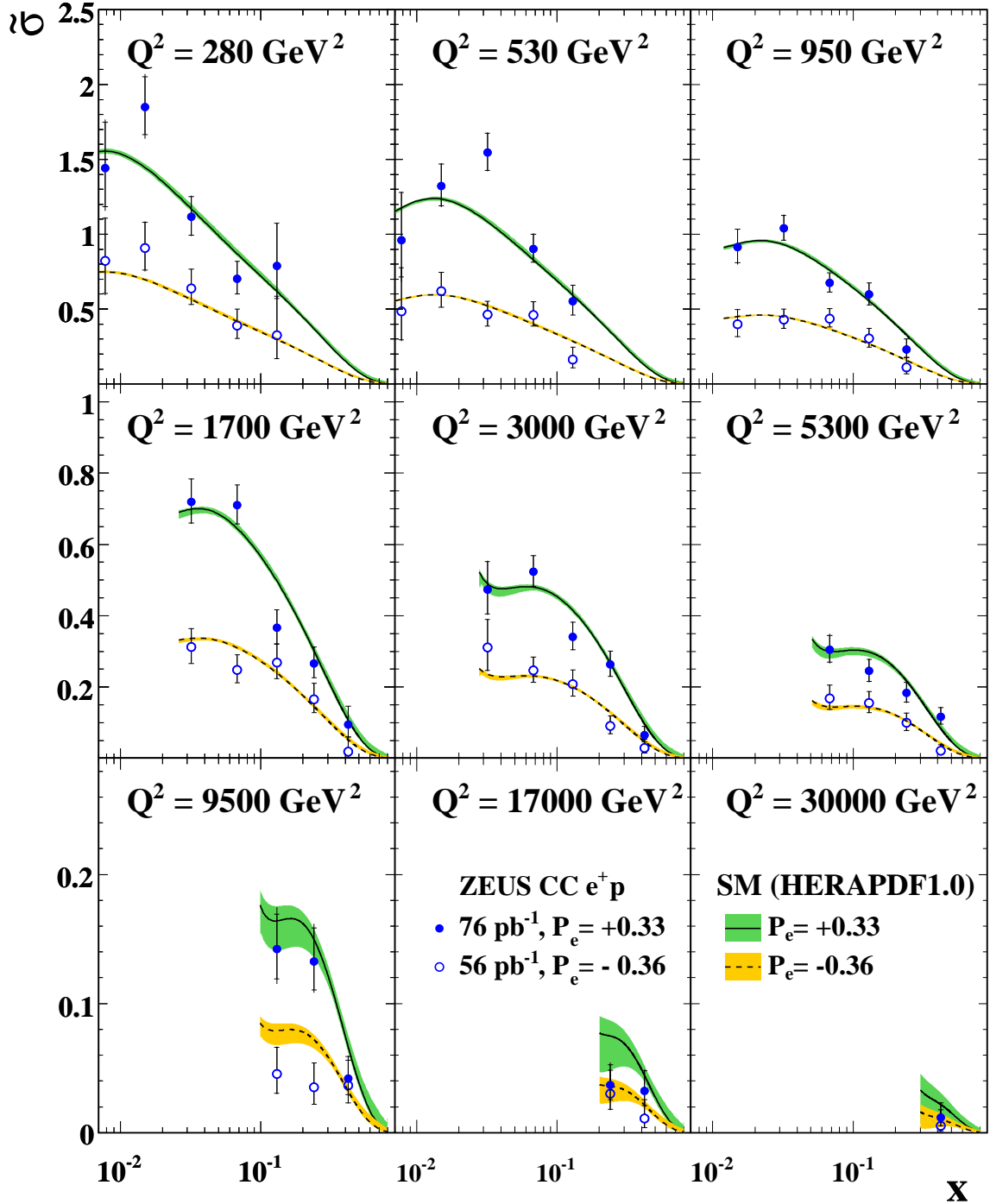}
  \end{center}
  \caption{ The $e^+p$ CC DIS reduced cross section plotted as a
    function of $x$ for fixed $Q^{2}$.  The positive (negative) 
    polarisation data are shown as the filled (open) points. 
    The curves show the predictions of the SM evaluated using the HERAPDF1.0
    PDFs. The shaded bands show the total uncertainty from the 
    HERAPDF1.0 PDFs.}
  \label{fig-double-separate}
\end{figure}

\begin{figure}
  \begin{center}
    \includegraphics[width=1.0\textwidth]{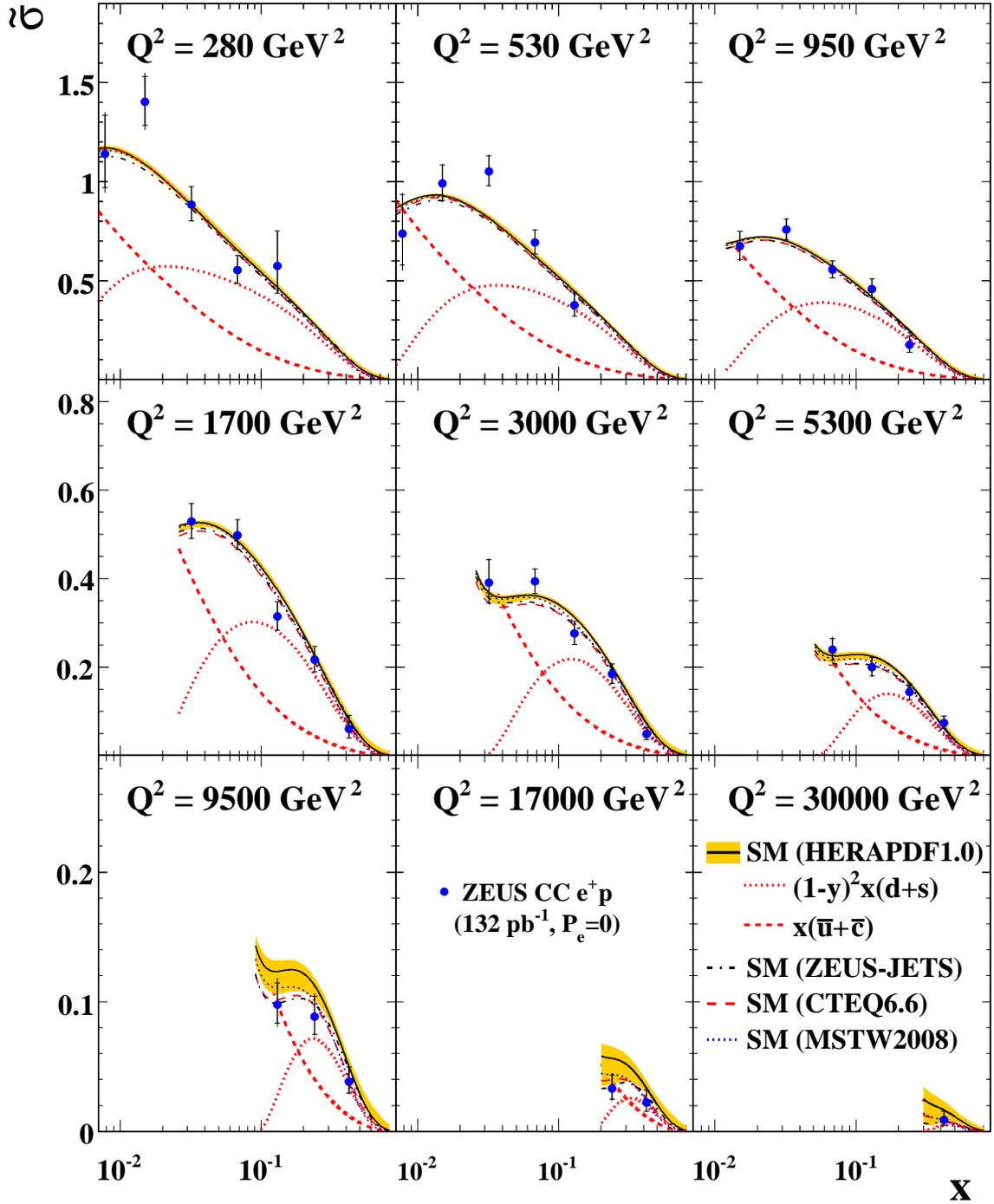}
  \end{center}
  \caption{ The $e^+p$ CC DIS reduced cross section plotted as a
    function of $x$ for fixed $Q^{2}$.   The circles represent the
    data points and the curves show the predictions of the SM
    evaluated using the HERAPDF1.0, ZEUS-JETS, CTEQ6.6 and MSTW2008 PDFs. The
    dashed and dotted lines show the contributions of the PDF
    combinations $(1-y)^{2}x(d+s)$ and $x(\bar{u}+\bar{c})$,
    respectively.  }
  \label{fig-double}
\end{figure}

\begin{figure}
  \begin{center}
    \includegraphics[width=1.0\textwidth]{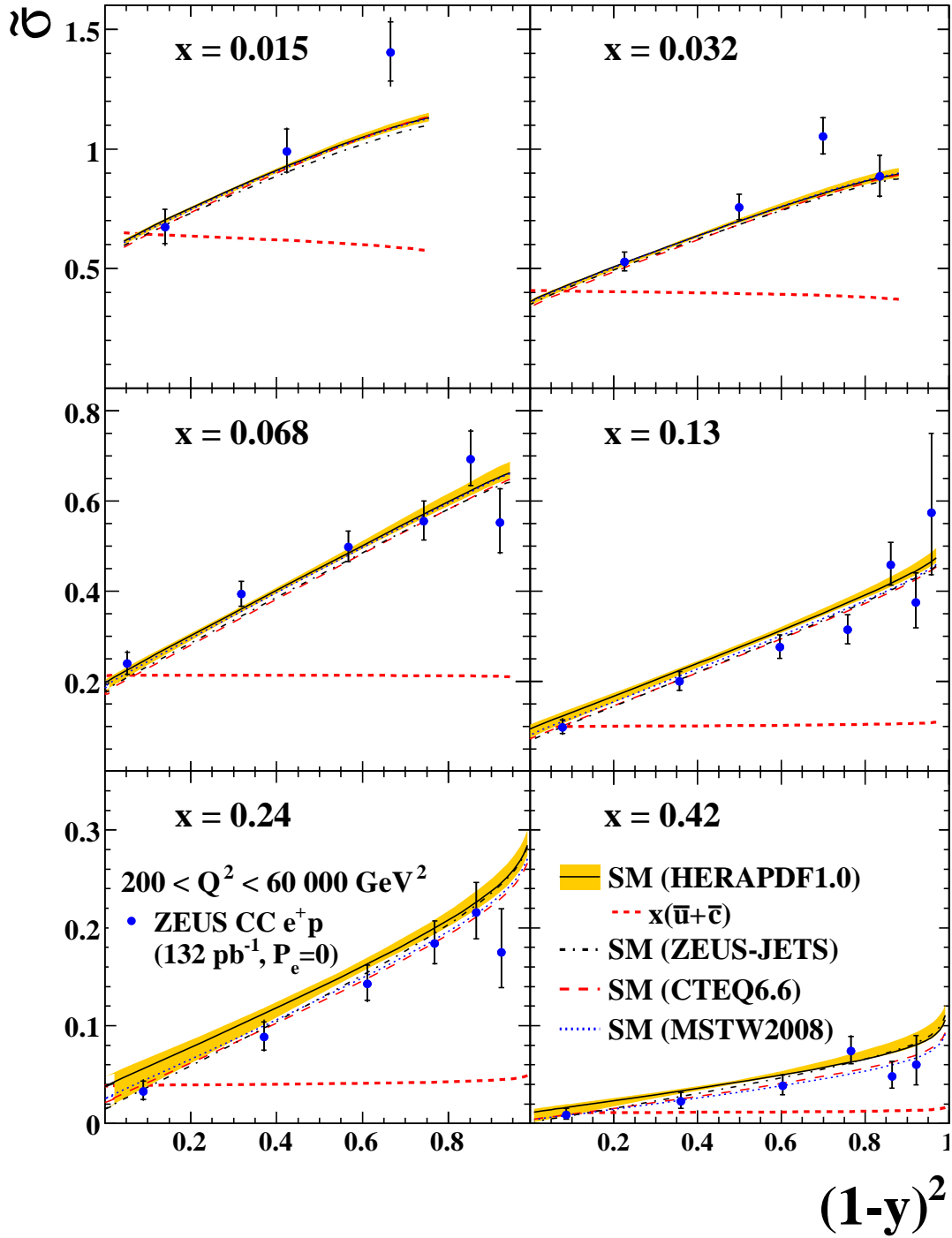}
  \end{center}
  \caption{ The $e^+p$ CC DIS reduced cross section plotted as a
    function of $(1-y)^2$  for fixed $x$. The circles represent the
    data points and the curves show the predictions of the SM
    evaluated using the HERAPDF1.0, ZEUS-JETS, CTEQ6.6 and MSTW2008 PDFs. The
    dashed lines show the contributions of the PDF  combination
    $x(\bar{u}+\bar{c})$ and the shaded band shows the total  uncertainty
    from the HERAPDF1.0 PDFs.}
  \label{fig-helicity}
\end{figure}

\begin{figure}[p]
  \begin{center}
    \includegraphics[width=.8\textwidth]{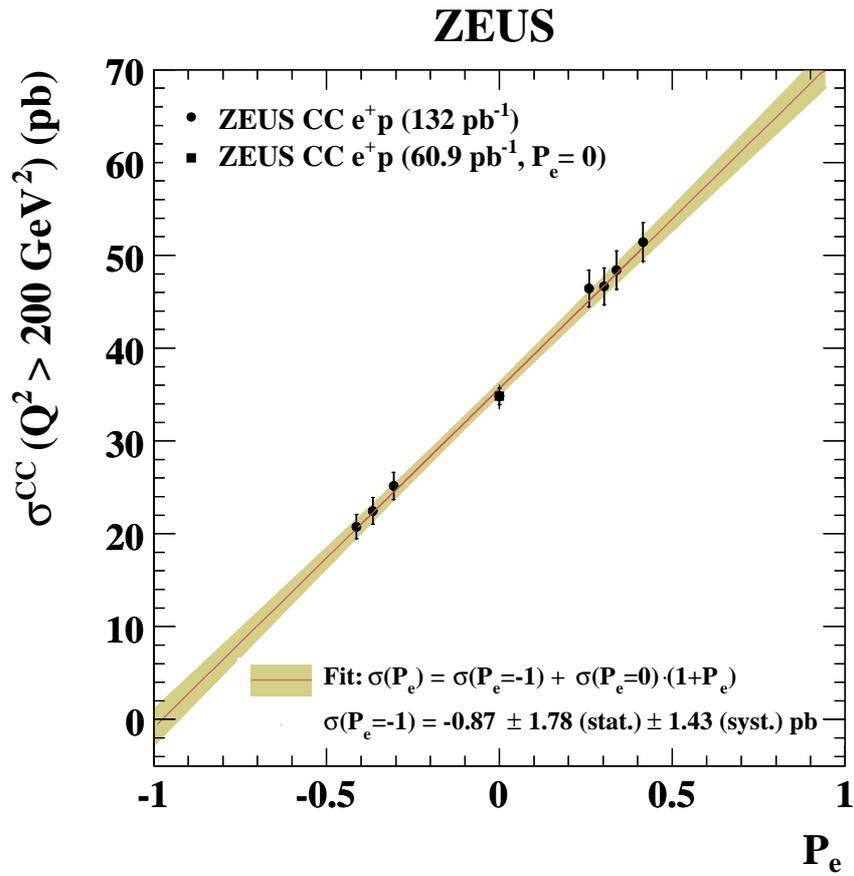}
  \end{center}
  \caption{The total cross sections for $e^+ p$ CC DIS as a function
    of the longitudinal  polarisation of the positron beam. The line 
    shows the linear fit to the points and the shaded band shows the 
    uncertainty of the fit.} 
\label{fig-polbins}
\end{figure}